\documentclass[aps,prb,twocolumn,superscriptaddress]{revtex4-2}

\usepackage{dcolumn}% Align table columns on decimal point
\usepackage{bm}% bold math

\usepackage[utf8]{inputenc}
\usepackage[english]{babel}
\usepackage{subcaption}
\usepackage{graphicx}
\usepackage{wrapfig}
\usepackage{amsmath}
\usepackage{float}
\usepackage{amssymb}
\usepackage{color}
\usepackage[normalem]{ulem}
\usepackage{tikz}
\usepackage{multirow} 
\usepackage{array}

\usepackage{xcolor}

\addto\captionsenglish{}
\addto\captionsenglish{}

\newcommand{\liq}{{\rm liq}}
\newcommand{\sol}{{\rm sol}}

\begin{document}

%\title{A physics-informed AI method for calculating melting points with uncertainty control and optimal sampling}
\title{Accurate Melting Point Prediction through Autonomous Physics-Informed Learning}

\author{Olga Klimanova}
\affiliation{Moscow Institute of Physics and Technology, Russia}
\affiliation{Skolkovo Institute of Science and Technology, Russia}

\author{Timofei Miryashkin}
\affiliation{Skolkovo Institute of Science and Technology, Russia}

\author{Alexander Shapeev}
\affiliation{Skolkovo Institute of Science and Technology, Russia}

\date{\today}

\begin{abstract}
We present an algorithm for computing melting points by autonomously learning from coexistence simulations in the NPT ensemble. Given the interatomic interaction model, the method makes decisions regarding the number of atoms and temperature at which to conduct simulations, and based on the collected data predicts the melting point along with the uncertainty, which can be systematically improved with more data. We demonstrate how incorporating physical models of the solid-liquid coexistence evolution enhances the algorithm's accuracy and enables optimal decision-making to effectively reduce predictive uncertainty. To validate our approach, we compare the results of 20 melting point calculations from the literature to the results of our calculations, all conducted with same interatomic potentials. Remarkably, we observe significant deviations in about one-third of the cases, underscoring the need for accurate and reliable algorithms for materials property calculations.
\end{abstract}

\maketitle

\section{Introduction}

With the constantly increasing computational power of modern computers, the demand in efficient prediction of materials properties and computationally generating new materials grows. One such property is a phase diagram, which contains information on which phases are thermodynamically stable at a given temperature, composition, and pressure.

The CALPHAD method represents the state-of-the-art approach in constructing phase diagrams for practical problems \cite{Saunders1998,SPENCER2008}. CALPHAD applies a classical fitting approach to represent the Gibbs free energy functions of different phases. The free energy is primarily fitted to the experimental data, thus resulting in an empirically derived phase diagram.

Nevertheless, the methods of constructing phase diagrams without the extensive use of experimental data (in this sense, from first principles) are rapidly maturing. If we know the model according to which the atoms interact, then we can run molecular dynamics (MD), calculate averages of quantities such as potential energy and pressure over the MD trajectories, and use these data to reconstruct the free energy as a function of thermodynamic variables (such as temperature, composition, or pressure). This method is called thermodynamic integration \cite{Frenkel} as the data used to reconstruct the free energy are the derivatives of the free energy with respect to the thermodynamic variables. Then, by minimizing the free energy, the phase diagram can be constructed from the free energy functions of each phase.

A key component to the practical use of the thermodynamic integration method for predictive materials modeling is machine-learning potentials \cite{behler_generalized_2007,bartok_gaussian_2010,thompson_spectral_2015,shapeev_moment_2016,drautz_atomic_2019,jinnouchi2020-gaussian,wang2018-deepmd,pun2020-pinn,smith2021-ani-al,batzner_e3-equivariant_2022}.
Such potentials are trained on a data set of atomic configurations with corresponding energies, forces, and stresses obtained from the density functional theory (DFT) calculations. In this work, in addition to adopting conventional interatomic potentials from the literature, we employ MTPs (moment tensor potentials) that we train in an automated manner following the approach outlined in \cite{Novikov}.

One difficulty of the thermodynamic integration method is that the free energy of phases is accurately reconstructed only up to a constant---indeed, derivatives of the free energy do not contain information about its additive shift. And even though only relative shifts of the free energy of different phases matter, additional data containing such information need to be incorporated into the thermodynamic integration \cite{Cheng2021} or Bayesian reconstruction methods \cite{Ladygin}. When reconstructing the free energy of a solid phase that is stable at $T=0$, one can use the Hessian (or force constant) information, which provides the absolute value of the free energy limit as $T\to 0$. However, if we are also interested in the liquid phase, we need to compute melting points to determine the difference in the free energy of the solid and liquid phases. Thus, a method for calculating melting points is an important part of computational protocols for phase diagram calculation. Moreover, if such a method was to be used with a Bayesian reconstruction method \cite{Ladygin}, it is necessary to predict the melting point with reliable uncertainties (confidence intervals).

At the moment, there are many ways to calculate the melting point: the interface pinning method \cite{Pedersen2013}, the hysteresis method \cite{Luo2004,Zheng2007}, and the two-phase coexistence method. The latter comes in different flavors depending on the ensemble used, such as NVE \cite{Morris}, NPH \cite{WangNPH}, and NPT \cite{Walle}.

The approach most relevant to our work was the one introduced by Hong and van de Walle \cite{Walle}, in which the method of coexistence in the NPT ensemble was used. In this method, an initial atomic configuration consists of two roughly equal parts: solid and liquid. Then, MD is run during which the solid/liquid interface randomly moves. The simulation temperature may not exactly coincide with the melting temperature, resulting in a drift towards either the solid or liquid phase. The MD is continued until the configuration either fully solidifies or melts. The collected data include the conditions of the MD (number of atoms, temperature, and pressure) and the binary outcome of the MD (``solid'' or ``liquid''). The melting point is determined from these data. The authors of \cite{Walle} propose a statistical method for postprocessing the results of the MD runs. This method is based on a statistical-physics model that describes how the solid-liquid interface moves depending on the difference in free energies of the two phases at a given temperature. The unknown parameters are fitted from the MD data, and the melting point is obtained as the temperature at which the probability of ending up in the solid or liquid phase is both equal to 1/2.

In the present work, we propose a new algorithm capable of autonomously calculating the melting point of materials based on Bayesian learning and sampling and the NPT-coexistence method. The method consists of two levels: at the first level, the melting point is predicted for the fixed number of atoms by running MD simulations at different temperatures (and possibly pressures), while at the second level the data for different number of atoms $N$ are combined to find the limit of the melting point as $N\to\infty$. At the first level, our method is, in part, equivalent to the method of Hong and van de Walle \cite{Walle}---it essentially gives the same formula for the melting point at the fixed number of atoms. However, instead of simply fitting the parameters of the formula to the data, we use the Bayesian approach which provides us with the statistical error (i.e., uncertainty) which is very important for the downstream tasks like phase diagram construction using Bayesian regression \cite{Ladygin}. At the second level, conventional Gaussian processes are used to postprocess the results of the melting point calculations for different number of atoms and hence evaluate the convergence of the melting point as a function of the number of atoms in the simulation, as well as to extrapolate the result to an infinite number of atoms, thus evaluating not only the statistical error but also eliminating the systematic error associated with the finite number of atoms. At both levels, the Bayesian framework allows for choosing the parameters ($N$ and $T$) for conducting MD calculations in order to reduce the uncertainty in the final answer in an optimal manner, effectively automating the decision-making process typically carried out by a researcher.

The present manuscript is organized as follows. In Section \ref{sec:methods}, we present our melting point calculation method. We emphasize how we utilize physics to inform our algorithm for making predictions and decisions regarding the calculation of the melting point. Specifically, we solve the Fokker-Planck equations that describe the motion of the solid-liquid interface. This enables us to derive the nonlinear Bayesian regression likelihood formula for the melting point and determine the model hyperparameters. In Section \ref{sec:results}, we present and discuss the results obtained using our method. We compare approximately 20 results from the literature with those obtained with our method all calculated with the same interatomic potentials. Interestingly, in a significant number of cases, we found discrepancies that cannot be explained by shortcomings of existing methods, such as finite-size effects, or by possible underprediction of the melting point error by our method. Finally, in Section \ref{sec:conclusions}, we provide concluding remarks.

\section{The Melting Point Calculation Method}\label{sec:methods}

The core of our methodology is Bayesian regression. We present the methodology for zero-pressure calculations, although we emphasize that it is generalizable to arbitrary-pressure calculations. Furthermore, as mentioned earlier, the problem of melting point calculation is separated into predicting the melting point $T^*$ for a given number of atoms, $N$, at the first level, and then reconstructing the dependence of $T^*$ on $N$ at the second level.
Thus, at the first level, i.e., for the fixed $N$, we work with data of the form $(T^{(i)}, n_{\rm s}^{(i)}, n_{\rm l}^{(i)})_{i=1,2,\ldots}$, where $T^{(i)}$ is the simulation temperature and $n_{\rm s}^{(i)}$ and $n_{\rm l}^{(i)}$ are the number of ``solid'' and ``liquid'' outcomes of the simulations, respectively.

The idea behind the Bayesian regression is that the probability density of $T^*$ being the melting point of the underlying system given the data, $p(T^*|T^{(i)}, n_{\rm s}^{(i)}, n_{\rm l}^{(i)})$, can be expressed through the reverse probability density, $p(T^{(i)}, n_{\rm s}^{(i)}, n_{\rm l}^{(i)}|T^*)$. The latter expresses how probable it is to observe the given data when the true melting temperature is $T^*$.
This requires us to construct the corresponding model, which in the case of linear dependencies results in prescribing the Gaussian process kernel, however, in our case it requires a more complex model which we will begin to derive in the next subsection.

\subsection{Model of Solid-Liquid Interface Motion} \label{sec:Model}

\begin{figure}[b]
\includegraphics[width=6cm]{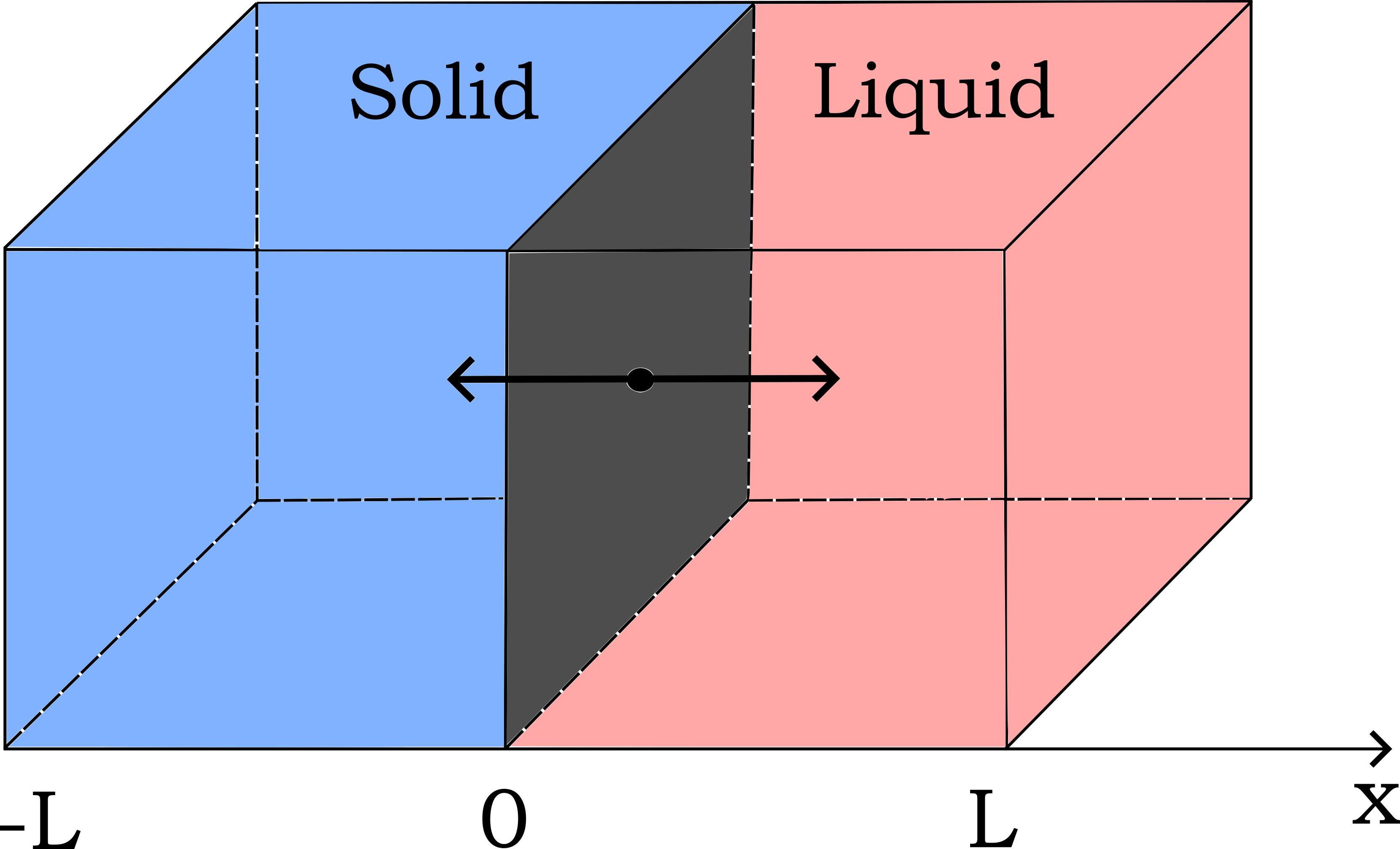} \\
\caption{\label{fig:solid-liquid-sketch} An illustration of one-dimensional solid-liquid interface motion. The interface starts at the position $x = 0$ and drifts towards the dominant phase during MD. The motion process is terminated when the interface reaches $x = \pm L$.
In the actual simulation we use the periodic boundary conditions, therefore there is another interface at $x = \pm L$ which is not shown in this sketch.
}
\end{figure}

We assume that the solid-liquid interface performs an essentially one-dimensional motion as illustrated in Figure \ref{fig:solid-liquid-sketch} and hence the position of the interface can be described by simply $x=x(t)$, where $t$ is time.
The interface starts at the position $x = 0$, randomly shifts right and left in the process of molecular dynamics and drifts towards the dominant (at a given temperature) phase.
Mathematically, we formulate this stochastic process as  
\begin{equation}\label{eq:proc_original}
    dx = \alpha dt + \sigma dW(t),
\end{equation}
with the drift coefficient $\alpha$ and diffusion coefficient $D = \sigma^2/2$, where $W(t)$ is the white noise. 
The corresponding Fokker-Planck equation for the probability density $p(x,t)$ of the random variable $x$ is
\begin{equation}\label{eq:proc_FP}
    \frac{\partial p(x,t)}{\partial t} = \bigg[-\alpha\frac{\partial}{\partial x} + D \frac{\partial^2}{\partial x^2}\bigg]p(x,t).
\end{equation}
If the interface $x(t)$ reaches $x = \pm L$ then the process is terminated.
The corresponding boundary conditions are hence
\begin{equation}\label{eq:proc_FP_bc}
p(-L,t) = p(L,t) = 0.
\end{equation}

The drift $\alpha$ depends on the temperature and we simply approximate it with a linear dependence assuming that higher-order Taylor expansion terms can be neglected near the melting point $T^*$:
\[
\alpha = -\gamma (T-T^*),
\]
where $\gamma$ is the first Taylor coefficient.
The diffusion coefficient $D$ does not depend on the temperature (at least when we are close to the melting point), but depends on the system size.
We assume that locally, around each atom at the interface, the system can randomly shift towards the solid or liquid phase.
Hence the contribution of one atom to the motion of the interface is ${\mathcal N}(0,(1/L^2)^2)$, in other words within a certain unit time interval it can move the interface by $O(L^{-2})$ (because there are $O(L^2)$ atoms on the interface, to move the interface by $O(1)$ all $O(L^2)$ need to move in the same direction).
All the atoms on the interface contribute the sum of $L^2$ of such random variables which is
\[
{\mathcal N}\big(0,L^2 (L^{-2})^2\big) = {\mathcal N}(0,L^{-2}).
\]
We thus assume that $D = \kappa L^{-2}$ with some coefficient $\kappa$ and we thus have the following system
\begin{align} \label{eq:FP_eq}
    \frac{\partial p(x,t)}{\partial t} &= \Bigl[\gamma (T-T^*) \frac{\partial}{\partial x} + \kappa L^{-2} \frac{\partial^2}{\partial x^2}\Bigr]p(x,t),
    \\ \label{eq:FP_bc}
    p(-L,t) &= p(L,t) = 0.
\end{align}

The probability of a process described by \eqref{eq:FP_eq} and \eqref{eq:FP_bc} to eventually reach the all-solid state is
\begin{equation}\label{eq:p_sol_formula}
p_\sol = \frac{1}{1 + e^{(T-T^*)/\sigma^*}},
\end{equation}
where we define $\sigma^* = \kappa/(\gamma L^3)$ and call it the \emph{melting temperature spread}.
We note that we do not independently find $\kappa$ and and $\gamma$ from data, but we indeed find $\sigma^*$ from the MD data that we collect.
From \eqref{eq:p_sol_formula} the probability of reaching the all-liquid phase follows:
\begin{equation}\label{eq:p_liq_formula}
    p_\liq = 1 - p_\sol = \frac{e^{(T-T^*)/\sigma^*}}{1 + e^{(T-T^*)/\sigma^*}}.
\end{equation}

We note that in the actual simulation we have periodic boundary conditions and therefore two (not one) interfaces between solid and liquid that move independently.
However, because we anyway do not determine $\kappa$ and $\gamma$ independently as a function of $L$, but only the combination $\sigma^* = \kappa/(\gamma L^3)$ from data, the derivation with the two simultaneously moving interfaces would be more involved, but result in the same formula as \eqref{eq:p_liq_formula}.
In the Results section below, Figure \ref{fig:pliq} displays the actual fit of \eqref{eq:p_liq_formula} along with the collected data.

\subsection{Dependence of the Simulation Time on $L$}\label{sec:simulation_time}

To optimize our melting point calculation protocol, we need to find how the simulation time depends on the system size, i.e., $\tau_{\rm exit} \sim L^{\beta}$, where the power exponent $\beta$ is to be determined.

To that end, we return back to the non-stationary Fokker-Planck equation \eqref{eq:FP_eq}--\eqref{eq:FP_bc}.
The simulation temperatures $T$ will be chosen to be close to $T^*$, such that $\gamma (T-T^*) L^3/\kappa = \delta$ with some constant $\delta$, ensuring that the exponent in \eqref{eq:p_sol_formula} has the value ($e^\delta$) and remains independent as $L$ changes.
The resulting equations are as follows:
\begin{align*}
    \frac{\partial p(x,t)}{\partial t} &= \Bigl[\delta \kappa L^{-3} \frac{\partial}{\partial x} + \kappa L^{-2} \frac{\partial^2}{\partial x^2}\Bigr]p(x,t),
    \\
    p(-L,t) &= p(L,t) = 0.
\end{align*}

Next, we make a change of variables, substituting $p(x,t)$ with $\tilde{p}(\tilde{x},\tilde{t})$, where $\tilde{x} = L^{-1} x$, $\tilde{t} = L^{-\beta} t$ and $\tilde{p} = p$.
Consequently, we substitute $\tilde{p}(\tilde{x},\tilde{t}) = p(x,t)$ into the Fokker-Planck equation, resulting in:
\[
\frac{\partial \tilde{p}(\tilde{x},\tilde{t})}{\partial \tilde{t}}L^{-\beta} = L^{-4}\kappa \Bigl[\delta \frac{\partial}{\partial \tilde{x}} + \frac{\partial^2}{\partial \tilde{x}^2}\Bigr]\tilde{p}(\tilde{x},\tilde{t}).
\]
We require that $\tilde{p}(\tilde{x},\tilde{t})$ does not depend on $L$, yielding $\beta = 4$. Hence, we find that $\tau_{\rm exit} = L^{4} \tilde{t}_{\rm exit}$.

The actual computation (CPU) time for a molecular dynamics simulation will then scale as $L^7$ (because we have $O(L^3)$ atoms).

\subsection{Nonlinear Bayesian regression} \label{sec:NonlinearBayes}

To calculate the melting point, we conduct MD simulations in an isothermo-isobaric (NPT) ensemble. To determine whether our two-phase system is melted or solidified, we compare the potential energy during the simulation with the defined \emph{solid threshold energy} and \emph{liquid threshold energy}, which will be specified below.

In coexistence simulations, an initial estimation of the system's melting point ($T_{\rm m}$) is required to facilitate the preparation of liquid and its coexistence with solid at a specific temperature. It is worth noting that in various material categories, such as metallic alloys, the melting point of a new material can be approximated using Vegard's law \cite{kittel1996introduction} or modern machine-learning based models \cite{hong2022melting}. In the general case, we defer the task of automatically estimating the melting point to future research. For the sake of reproducibility, we select the experimental melting point as the initial estimation for $T_{\rm m}$ in our work.

\begin{figure}[ht]
\begin{center}
\includegraphics[width=9.5cm]{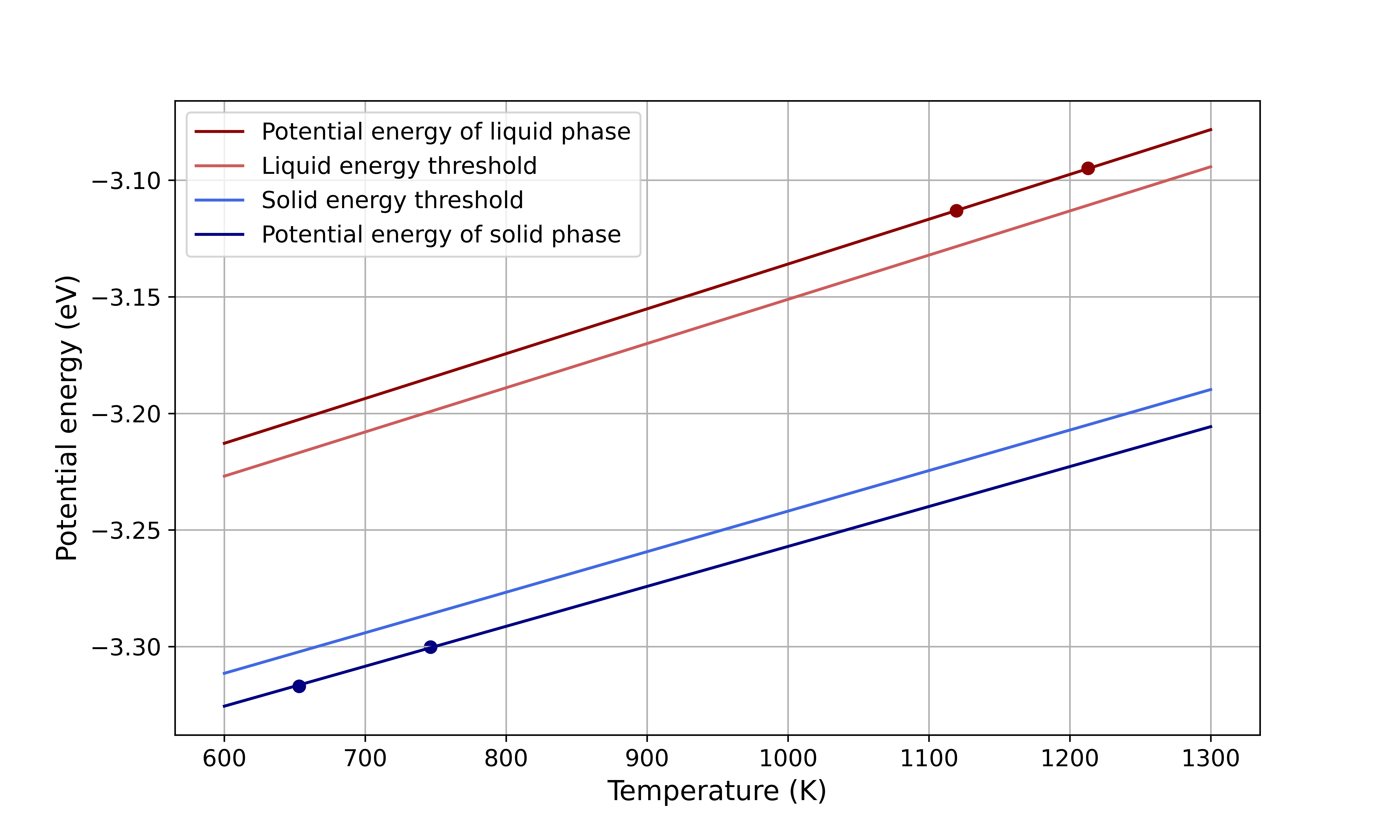}
\caption{\label{fig:epot} Dependence of potential and threshold energies on temperature. The solid threshold energy is chosen slightly higher than the average potential energy of the solid phase, while the liquid threshold is chosen slightly lower than the average liquid potential energy.}
\end{center}
\end{figure}

We next run MD simulations of the solid phase at two temperatures ($0.7 T_{\rm m}$ and $0.8 T_{\rm m}$), let the system equilibrate and calculate the average potential energy at these temperatures.
We then construct a linear dependence of the potential energy on temperature as illustrated in Figure \ref{fig:epot}.
Similarly, to compute the potential energy of the liquid phase, we obtain the average potential energies at two higher temperatures, $1.2\,T_{\rm m}$ and $1.3\,T_{\rm m}$ and construct the linear dependence of the liquid potential energy.
We then calculate the solid threshold by adding to the solid potential energy one-eighth of the differences between the liquid and solid energies.
Likewise, the liquid threshold is one-eighth lower than the liquid potential energy.
These thresholds are also illustrated in Figure \ref{fig:epot}.
The value of one-eighth should be considered a hyperparameter in our method. As will see in simulations (Figure \ref{fig:epot}), choosing a lower threshold would likely result in slightly longer trajectories, but the outcome will remain the same. Conversely, selecting a higher value may lead to minor changes in the outcomes of the coexistence simulations; however, this effect should decrease with the increase of the system size $L$ and eventually vanish in the limit $L\to\infty$, which is considered in Section \ref{sec:GP}.
%The reason we do not take the exact solid (or liquid) energy as the threshold is because we do not want to wait for a long time till the system fully solidifies---indeed, as we will see from the results (Figure \ref{fig:lines_time_en}) once the system's potential energy reaches a certain point (higher than the chosen solid threshold), it then solidifies very quickly and has a negligible chance of melting instead.

We are interested in the likelihood that $T, \sigma$ is the melting point of the system and its spread given data, and, thanks to the Bayesian approach, we can obtain it from the probability (density) of observing the given $T$ and $\sigma$:
\begin{equation}\label{eq:bayes}
    p(T, \sigma\vert \text{data}) = \frac{p(\text{data}\vert T, \sigma)\,p(T)\,p(\sigma)}{p(\text{data})}.
\end{equation}
The main part of the right-hand side is $p(\text{data}\vert T, \sigma)$.
To find it, we note that since all MD simulations are independent we simply multiply the probabilities \eqref{eq:p_sol_formula} and \eqref{eq:p_liq_formula} for each of the respective outcomes in our data:
\begin{equation}\label{eq:data_prob_density}
    p(\text{data}\vert T, \sigma) = \frac{\prod\limits_{i = 1}^n \exp \left({n_{\rm l}^{(i)}\cdot\frac{T^{(i)} - T}{\sigma}}\right)}{\prod\limits_{i = 1}^n \big(1 + \exp\left({\frac{T^{(i)} - T}{\sigma}}\big)\right)^{n_{\rm l}^{(i)} + n_{\rm s}^{(i)}}}.
\end{equation}

In \eqref{eq:bayes} we simply assume $p(T)={\rm const}$ and $p({\rm data})={\rm const}$, meaning that all melting temperatures and data distributions have equal probabilities.
For $p(\sigma)$ we, however, choose $p(\sigma) = \sigma^{-2}$, indicating a higher probability for smaller $\sigma$ compared to larger $\sigma$.

Then we can express the mean value for the melting point and its variance as
\begin{align} \label{eq:T_melt}
T^* &= \overline{T} =  \frac{\iint T\cdot p(T,\sigma\vert \text{data})\,{\rm d}T\,\sigma^{-2}\,{\rm d}\sigma}{\iint p(T,\sigma\vert \text{data})\,{\rm d}T \,\sigma^{-2}\,{\rm d}\sigma},
\\ \label{eq:var_Tmelt}
(\Delta T^*)^2 &= 
{\rm Var}(T) = 
\frac{\iint (T- T^*)^2\cdot p(T,\sigma\vert \text{data})\,{\rm d}T \, \sigma^{-2} \, {\rm d}\sigma}{\iint p(T,\sigma\vert \text{data}) \, {\rm d} T \, \sigma^{-2} \, {\rm d}\sigma}.
\end{align}
We, by the way, can see that integrating over $\sigma^{-2} {\rm d}\sigma$ is equivalent to integrating over ${\rm d}(1/\sigma)$ which is used in our implementation.

\subsection{Optimal variance reduction for $T^*$ and $\sigma^*$} \label{sec:improvement}

Given the current approximation of $T^*$ and $\sigma^*$ based on the data, we can determine the most efficient temperature $T$ for conducting the next MD simulation with the help of Bayesian regression.
To simplify our computational protocol, instead of adaptively choosing the next point $T$, we ask the following question: given the distribution \eqref{eq:data_prob_density} with the exact $T^*$ and $\sigma^*$, at which point would the calculation reduce the error the most?
We can quickly find out that calculating at $T = T^*$ reduces the error in determining $T^*$ the most; however, it does not accurately determine $\sigma^*$.
Thus, we consider performing MD calculations at two temperatures, $T = T^* \pm \sigma^* x$, where the optimal value of unknown parameter $x$ will be determined shortly, and examine their effect on the variance of $\sigma^*$.

\begin{figure}[b]
\begin{center}
\includegraphics[width=9cm]{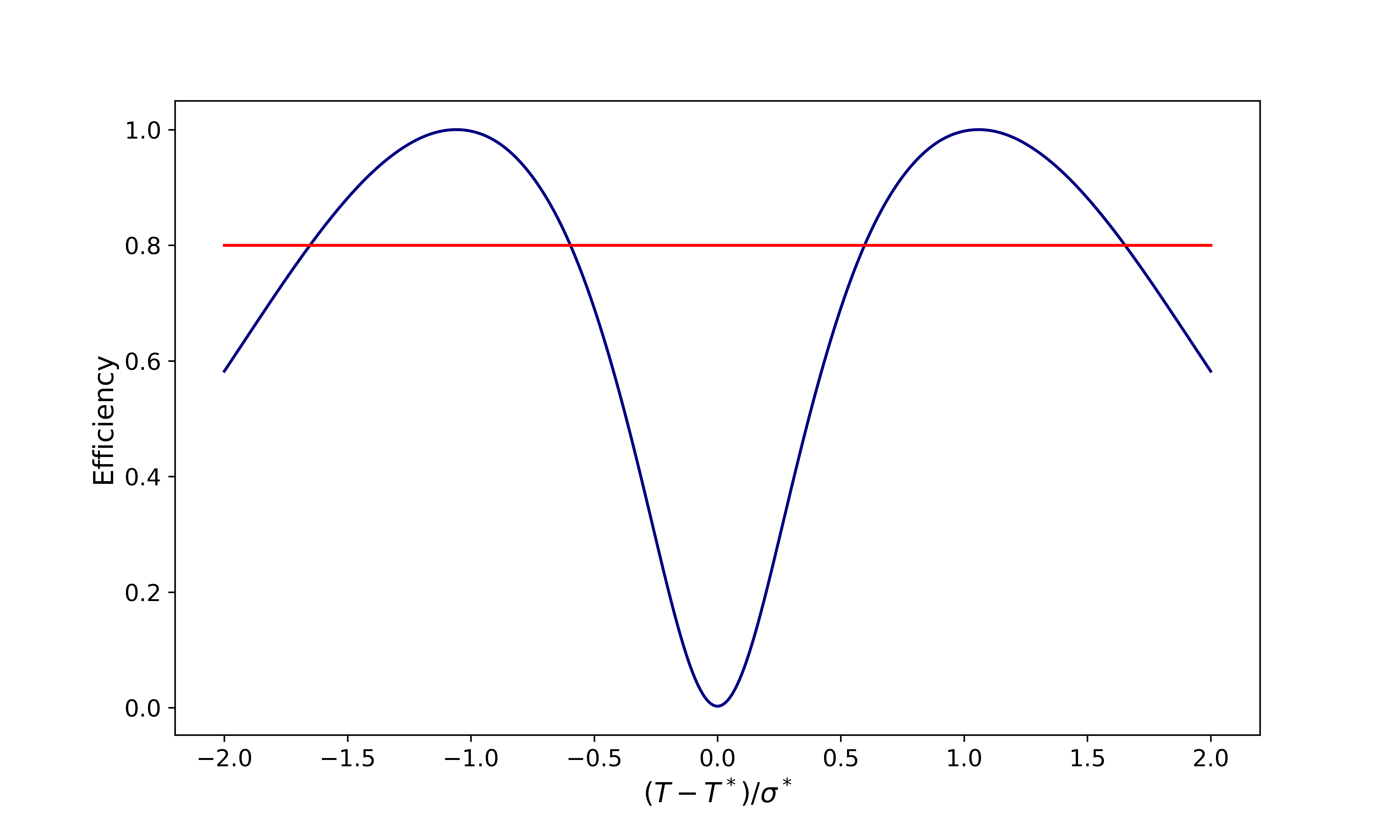}
\caption{\label{fig:efficiency}
Efficiency of running a simulation as a function of simulation temperature $T$.
Efficiency is understood as optimal reduction of uncertainty in predicting $\sigma^*$.
To achieve our target efficiency of at least 80\% (with 100\% corresponding to the best reduction of uncertainty), we choose the simulation temperatures as $T \in (T^* - 1.6\sigma^*, T^* - 0.6\sigma^*)$ and $T \in (T^* + 0.6\sigma^*, T^* + 1.6\sigma^*)$, where $T^*$ and $\sigma^*$ represent the Bayesian model predictions.
}
\end{center}
\end{figure}

We assume we perform a total of $n$ calculations at both temperatures, and the number of outcomes is proportional to their respective probabilities.
For liquid, at $T_{\rm md} = T^* + \sigma^* x$ the probability will be $e^x/(1+e^x)$ and at $T_{\rm md} = T^* - \sigma^* x$ it will be $1/(1+e^x)$.
We similarly compute the probabilities for solid, and after straightforward calculations, we find the following likelihood:
\begin{widetext}
\[
p(T,\sigma) = \frac{\exp\big( n (1+e^x)^{-1} \sigma^{-1} (T-T^* - \sigma^* x)\big) \exp\big( n (1+e^{-x})^{-1} \sigma^{-1} (T-T^* + \sigma^* x)\big)}{\big( 1 + \exp(\sigma^{-1} (T-T^* - \sigma^* x)) \big)^n \big( 1 + \exp(\sigma^{-1} (T-T^* + \sigma^* x)) \big)^n},
\]
\end{widetext}
where $T$ and $\sigma$ are the melting point and the distribution spread that we are trying to infer and $T^*$ and $\sigma^*$ are the true values.
Using this distribution, we find the variance in determining $\sigma^*$ as
\[
({\rm Var}(\sigma))^{1/n}
= \frac{%
\int_{-\infty}^{\infty} (\sigma-\sigma^*)^2 p(T,\sigma) {\rm d}T
}{%
\int_{-\infty}^{\infty} p(T,\sigma) {\rm d}T
}.
\]
Following \cite{Ladygin}, we think of the inverse variance-per-data point, $({\rm Var}(\sigma))^{-1/n}$ as (the amount of) information that the calculations at $T_{\rm md} = T^* \pm \sigma^* x$ give us about $\sigma$ (to be precise, this is the exponent of the information).
Such information as a function of $x$, relative to the maximum information (occurring at $x\approx \pm 1.06$) is plotted in Figure \ref{fig:efficiency}.
We do not choose the exact optimum in order to prevent too many different data points added to the training set.
Instead, we select $x$ from the intervals $[-1.6, -0.6]$ and $[0.6, 1.6]$, ensuring at least 80\% efficiency.
More precisely, we run MD simulations at $T \in (T^* - 1.6\sigma^*, T^* - 0.6\sigma^*)$ and $T \in (T^* + 0.6\sigma^*, T^* + 1.6\sigma^*)$, always trying to choose the point that already exists in the data set---this way we reduce the computational cost of our numerical integration.

\subsection{Gaussian process regression} \label{sec:GP}
The predicted melting temperatures $T^*_i$ and their uncertainties $\Delta T^*_i$ for different number of atoms $N_i$, are processed, at the second level, with the conventional Gaussian process regression.

Different numbers of atoms are collected in an input (feature) vector
\[
\boldsymbol{X} = \begin{pmatrix} 
     N_1 \\
     N_2 \\
     ... \\
     N_n
\end{pmatrix},
\]
while the vector of targets, together with their uncertainties are
\[
\boldsymbol{y} = \begin{pmatrix}
     T^*_1 \\
     T^*_2 \\
     ... \\
     T^*_n
\end{pmatrix}
,\quad
\boldsymbol{\Delta y} = \begin{pmatrix}
     \Delta T^*_1 \\
     \Delta T^*_2 \\
     ... \\
     \Delta T^*_n
\end{pmatrix}
,
\]
where $n$ is the number of observations.
We apply the Gaussian process regression \cite{Rasmussen} with the kernel
\begin{equation}\label{eq:kernel}
    k(N_1, N_2) = \theta^{2}_{f}\exp\biggl(-\Bigl(\frac{1}{N_1} - \frac{1}{N_2}\Bigr)^2\frac{\theta_N^2}{2}\biggr),
\end{equation}
with the hyperparameters $\theta_{f}$ and $\theta_{N}$ that are found from data.
This kernel expresses the expectation that the melting point converges as $O(1/N)$, but the speed of convergence ($\theta_N$) and the scale of the melting point itself ($\theta_f$) are unknown, and we obtain them from the data.
Importantly, we can legally substitute $N=\infty$ in the kernel which would tell us how melting point at a finite $N$ is correlated with that at infinite $N$ which is exactly what we are after.
In a slightly informal language, we can say that our algorithm ``understands'' and predicts the convergence of the melting point as $N\to\infty$.

%\begin{equation*}
%\begin{pmatrix}
%\boldsymbol{y}\\
%{y^*}
%\end{pmatrix}
%\sim
%\mathcal{N}\begin{pmatrix} \begin{pmatrix} \boldsymbol{0} \\ 0 %\end{pmatrix}, \begin{pmatrix} K(\boldsymbol{X}, %\boldsymbol{X}) + \rm diag(\boldsymbol{\Delta y^2}) & %K(\boldsymbol{X}, X_{*}) \\ K(X_{*}, \boldsymbol{X}) & K(X_{*}, %X_{*}) \end{pmatrix} \end{pmatrix}
%\end{equation*}
%Predictive mean:

Applying the Gaussian process machinery yields the mean for the predicted value $y^{*}$ at the point $X_{*}$ given by
\begin{equation*}
    \overline{y^{*}} = K(X_{*}, \boldsymbol{X})\Bigl[ K(\boldsymbol{X}, \boldsymbol{X}) + {\rm diag}(\bm{\Delta y^2}) \Bigr]^{-1}\bm{y},
\end{equation*}
and its variance
\begin{eqnarray*}
{\rm Var}(y^*) =&& K(X_{*}, X_{*}) - K(X_{*}, \boldsymbol{X})\nonumber\\
&&\cdot
\Bigl[ K(\boldsymbol{X}, \boldsymbol{X}) + {\rm diag}(\boldsymbol{\Delta y}^2) \Bigr]^{-1}K(\boldsymbol{X}, X_{*}),
\end{eqnarray*}
where $K(X_*, \boldsymbol{X})$ is a vector with entries $k(X_*, N_i)$, 
$K(\boldsymbol{X}, \boldsymbol{X})$ is a matrix with entries $k(N_i, N_j)$, and ${\rm diag}(\boldsymbol{\Delta y}^2)$ represents a diagonal matrix with entries $\Delta T_i^2$ on the diagonal.
As mentioned, we are interested in the melting point for the ``infinite'' number of atoms, so we let $X_{*} = \infty$ and consider the corresponding $\overline{y^{*}}$ and ${\rm Var}(y^*)$ as the end result of our modeling.

The hyperparameters $\theta_{f}$ and $\theta_{N}$ are found using the maximum likelihood principle \cite{Rasmussen}, i.e., by maximizing
\begin{eqnarray*}
{\rm log}p(\boldsymbol{y}| \boldsymbol{X}, \theta_{f}, \theta_{N}) = -&& \frac{1}{2}\boldsymbol{y}^T\Bigl[ K(\boldsymbol{X}, \boldsymbol{X})+ {\rm diag}(\boldsymbol{\Delta y^2})\Bigr]^{-1}\boldsymbol{y}\nonumber\\
-&& \frac{1}{2} {\rm log}\det\Bigl[ K(\boldsymbol{X}, \boldsymbol{X}) + {\rm diag}(\boldsymbol{\Delta y^2})\Bigr]\nonumber\\
-&& \frac{n}{2} {\rm log}(2\pi).
\end{eqnarray*}

\subsection{Optimal sampling of numbers of atoms} \label{sec:OptimalSampling}

We next formulate a strategy to determining for which number of atoms $N_i$ we should run the next simulation.
That is, we want to optimize the variance in predicting the melting point in the limit of infinite number of atoms, $y^{\infty}$. This variance is
\begin{eqnarray}\label{eq:Vyinf}
{\rm Var}(y^{\infty}) =&& K(\infty, \infty) - K(\infty, \boldsymbol{X})\Bigl[ K(\boldsymbol{X}, \boldsymbol{X})\nonumber\\
+&& {\rm diag}(\boldsymbol{\Delta y}^2)\Bigr]^{-1}K(\boldsymbol{X}, \infty).
\end{eqnarray}
We need to find which uncertainty $\Delta y_i$ should be decreased by running more MD simulations in order to decrease ${\rm Var}(y^{\infty})$ most efficiently. %, considering that the effort in conducting a new simulation scales as $L_i^7$.
In other words, we need to calculate for which $i$ the derivative
$\frac{\partial {\rm Var}(y^{\infty})}{\partial {\rm (efforts)}_i}$ has the largest negative value, where by ${\rm (efforts)}_i$ we mean the computational efforts required to run the MD simulation of the system of size $L_i$.

Let $n_{\rm tot}^{(i)} = n_{\rm s}^{(i)} + n_{\rm l}^{(i)}$ be the total number of MD simulations with $N_i$ atoms.
We have that $\Delta y_i \sim \sigma^*_i (n_{\rm tot}^{(i)})^{-1/2}$, where $\sigma^*_i$ is the melting temperature spread for $N_i$ atoms (the formula expresses the fact that the uncertainty in determining $y_i$ is proportional to the data spread $\sigma^*_i$ and decreases proportionally to the square root of the data size $n_{\rm tot}^{(i)}$).
Therefore $(\Delta y_i)^{-2} \sim (\sigma^*_i)^{-2} n_{\rm tot}^{(i)}$ and we can estimate
\[
\frac{{\rm d}\Delta y_i^{-2}}{{\rm d}n_{\rm tot}^{(i)}} \sim \big(\sigma_i^*\big)^{-2}.
\]
As we derived at the end of Section \ref{sec:GP}, the computational effort to run each of the $n_{\rm tot}^{(i)}$ MD simulations is proportional to $L_i^{7}$, hence we have that 
\[
\frac{{\rm d}\Delta y_i^{-2}}{{\rm d}{\rm (efforts)}_i} \sim L_i^{-7}\sigma_i^{*^{-2}}.
\]

Finally, we express
\[
\frac{\partial {\rm Var}(y^{\infty})}{\partial {\rm (efforts)}_i}
=
\frac{\partial {\rm Var}(y^{\infty})}{\partial (\Delta y_i)^{-2}}
\frac{\partial (\Delta y_i)^{-2}}{\partial {\rm (efforts)}_i},
\]
calculate $\frac{\partial {\rm Var}(y^{\infty})}{\partial (\Delta y_i)^{-2}}$ from \eqref{eq:Vyinf}, and we obtain the desired $\frac{\partial {\rm Var}(y^{\infty})}{\partial {\rm (efforts)}_i}$.

\section{Results and Discussion}\label{sec:results}

\begin{figure}[hb]
\begin{center}
\includegraphics[width=6cm]{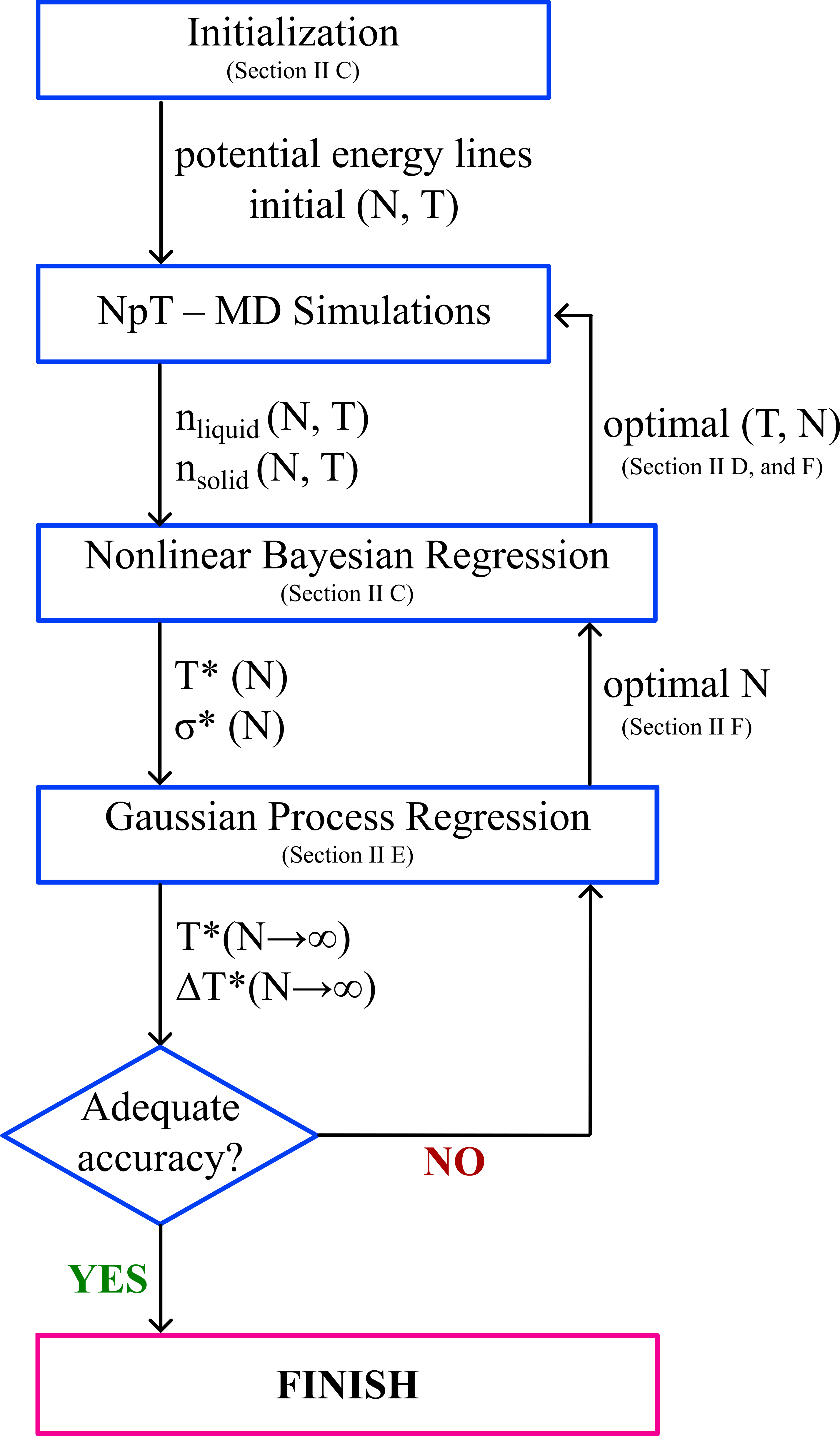}
\caption{\label{fig:diagram}
General scheme of our automatic method for calculating melting point.
We start with the dependence of the MD-averaged potential energies of solid and liquid as a function of temperature $T$ (``potential energy lines'').
We then perform MD simulations and process its data with our nonlinear Bayesian regression to obtain the melting point ($T^*$) and its spread ($\sigma^*$) as functions of the number of atoms $N$.
Then, with a Gaussian process, we extrapolate the results to the infinite system, and go back to MD simulations until we reach the desired accuracy.
In each cycle, with the help of the Gaussian process and Bayesian regression our algorithm decides which $N$ and $T$ to run the MD simulations with to reduce the uncertainty most efficiently.
} 
\end{center}
\end{figure}

We implemented the automatic method for the melting point calculations as described in the previous section.
We used the LAMMPS package \cite{LAMMPS} to simulate MD in the isothermo-isobaric (NPT) ensemble.
We tested our method on different interatomic potentials and crystal structures.
The general scheme of our approach is shown in Figure \ref{fig:diagram}.

\subsection{First example: Aluminum} \label{sec:aluminum}
In this section we present the results of application of our method for calculating the melting point of aluminum, while also discussing the functioning of our method.
To that end, we selected the Embedded Atom Method (EAM) potential proposed in \cite{Mendelev} as the potential describing aluminum.

The input data for our melting point (MP) calculations are the interatomic potential, crystal structure of the material, and desired accuracy of the MP calculations.
The algorithm first calculates the solid and liquid threshold energies, which are used as a termination criterion of molecular dynamics.
In order to calculate the solid threshold energy, our algorithm runs MD calculations at two temperatures, $0.7\,T_{\rm m}$ and $0.8\,T_{\rm m}$, where $T_m$ = 933 K for aluminum.
The calculations are run in the NPT ensemble for 15 ps and $8\times8\times16$ system size.
The algorithm averages the potential energy over the last 5 ps of each simulation and then reconstructs the solid threshold energy as a linear (with additive shift) function of the temperature.
To compute the liquid threshold energy, our algorithm obtains the average potential energies at two higher temperatures,
$1.2\,T_{\rm m}$ and $1.3\,T_{\rm m}$, with the same system size.
To make sure that the system melts, the algorithm first heats the system up at $1.5\,T_{\rm m}$ and keeps this temperature for 15 ps and then cools it down for 30 ps.
In the last 10 ps, the algorithm averages the potential energy and reconstructs the liquid threshold energy.

The algorithm next runs the coexistence MD simulations.
To make the first prediction of the MP, the algorithm prepares a  $3\times 3\times 6$ sample, i.e., $L=3$ or 216 atoms in total for the face-centered cubic (fcc) lattice --- we require a minimum of 200 atoms in the supercell.
%In the work \cite{Walle} the authors found out that the MP prediction error in such system is less than 100 K.
Then the algorithm takes $T_m$ as an initial guess, randomly chooses the value of $\sigma^* \in [10, 50]$ and finds two temperatures $T_1$ and $T_2$ as described in Section \ref{sec:improvement} to run 20 coexistence simulations (10 simulations at each temperature).

\begin{figure}[t]
\begin{center}
\includegraphics[width=9cm]{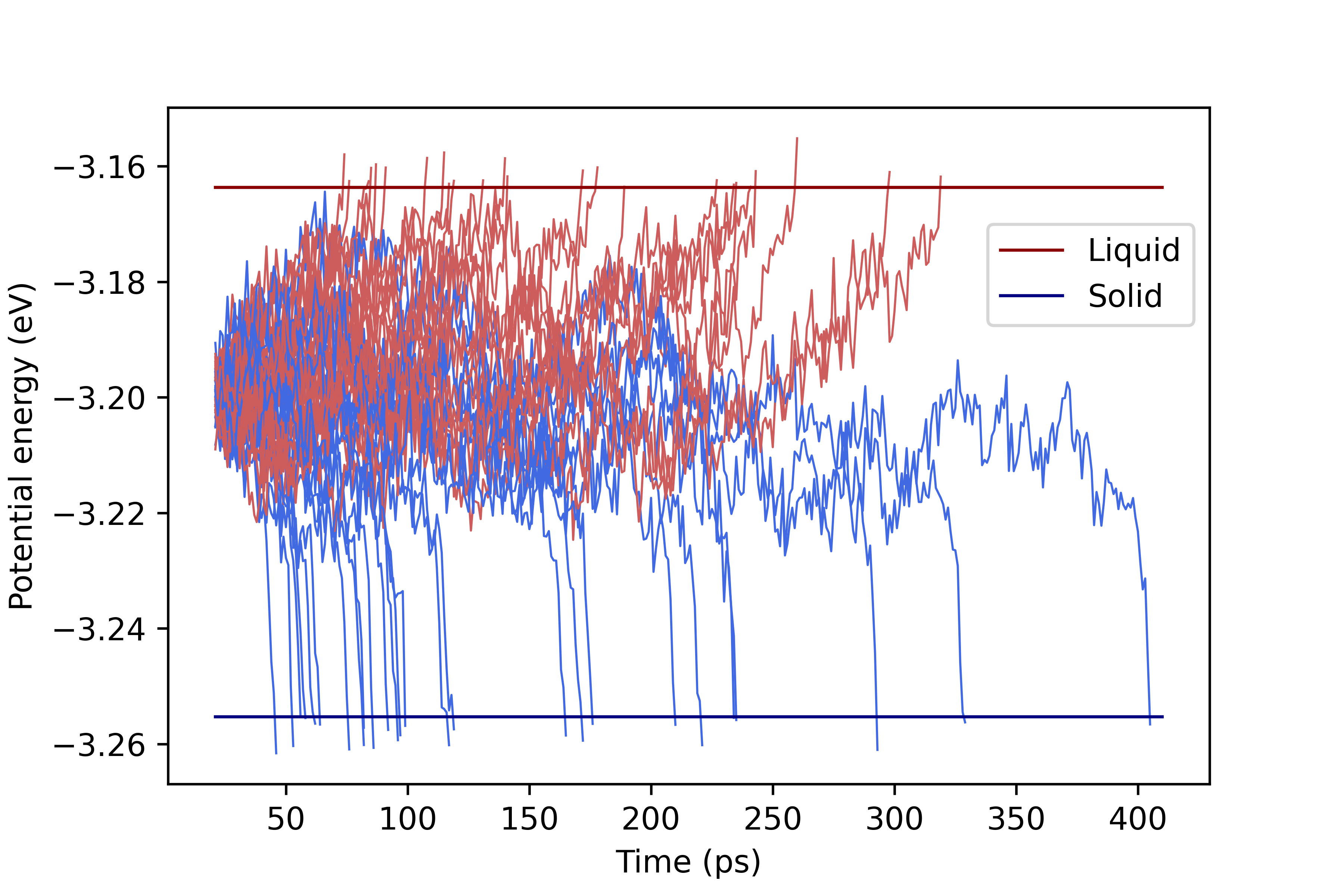}
\caption{Potential energy as a function of time for 50 independent MD runs for Al EAM ($6\times6\times12$ supercell, 1728 atoms) at 926 K. Each trajectory ends in a pure state, either solid (lower energy) or liquid (higher energy).}
\label{fig:lines_time_en} 
\end{center}
\end{figure}

The coexistence simulation is conducted as follows. Firstly, the algorithm performs an MD simulation at $T_1$ for 10 ps. At the end of this simulation, the system structure remains solid. Subsequently, half of the atoms are fixed (``frozen'') in their positions, while the other half is heated up to a temperature of $1.5\,T_{\rm m}$ for 10 ps.
After this MD phase, the system is prepared for the main coexistence MD run, which continues until the system's temperature reaches either of the two thresholds shown in Figure \ref{fig:lines_time_en}. 
The algorithm runs the coexistence simulation at temperature $T_2$ instead of $T_1$ in exactly the same way.

\begin{figure}[h]
\begin{center}
\includegraphics[width=8cm]{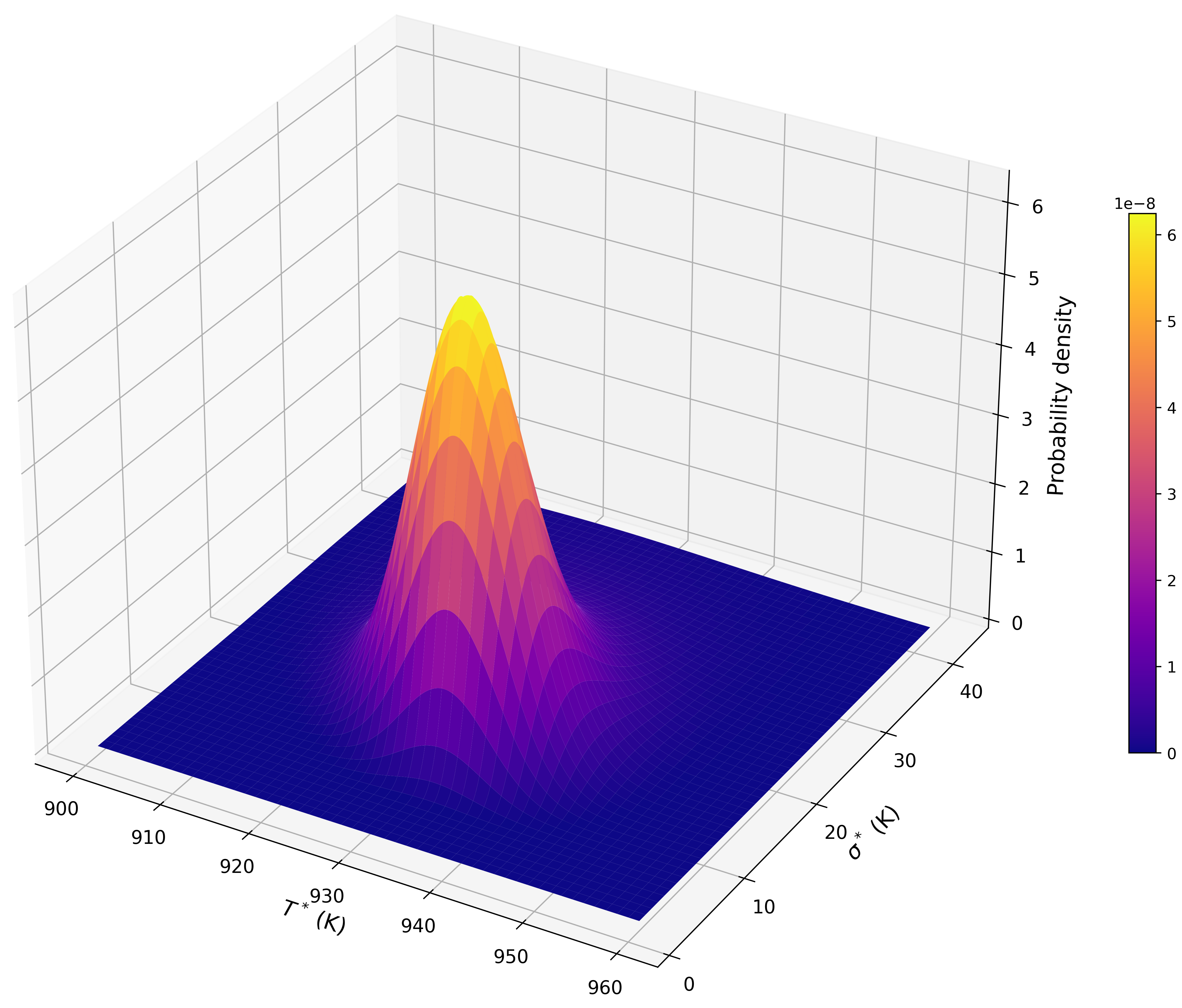}
\caption{Probability density of $T^*$ and $\sigma^*$ for the data given in Table \ref{tab:MD_data_AlEAM} for the aluminum EAM after two iterations of the algorithm.}
\label{fig:likelihood} 
\end{center}
\end{figure}

\begin{figure}[ht]
\begin{center}
\includegraphics[width=9.5cm]{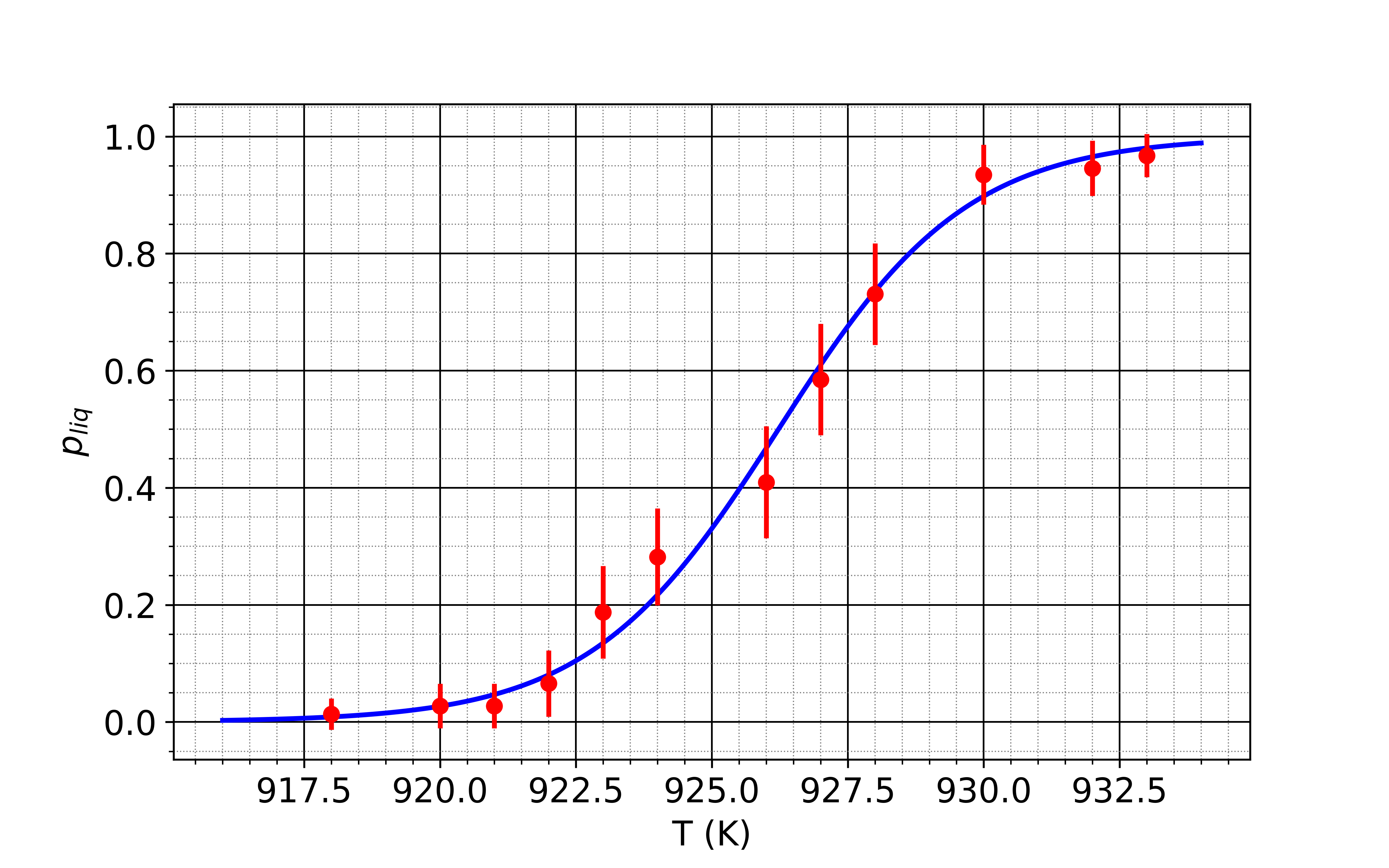}
\caption{\label{fig:pliq} Probability of reaching the all-liquid phase for aluminum at different temperatures for $L=6$. The fitted curve \eqref{eq:p_liq_formula} and the collected data with a 2-sigma confidence intervals are shown. All 12 data points agree with the fitted curve.}
\end{center}
\end{figure}

It is possible that all MD runs for the two temperatures yield the same outcome, either all solid or all liquid. In such cases, we adjust the simulation temperature, increasing or decreasing it, until we obtain both outcomes in our data for each system size $L$. Subsequently, we utilize the nonlinear Bayesian regression method described in Section \ref{sec:NonlinearBayes} to obtain the MP for at $L\in\{3,4,5,6\}$.
Throughout this process, the algorithm may identify new temperature values at which to conduct simulations. We provide an example of the data collected by the algorithm in Table \ref{tab:MD_data_AlEAM} and the posterior likelihood (probability density) for $T^*$ and $\sigma^*$ is plotted in Figure \ref{fig:likelihood}.
Also, for illustrative purposes we collected more data manually for $L=6$ and plotted the probability of reaching the liquid state \eqref{eq:p_liq_formula} as a function of temperature in Figure \ref{fig:pliq}.

\begin{table}[h]
\caption{\label{tab:MD_data_AlEAM} MD data for aluminum EAM fcc ($3\times3\times6$) after two iterations of our algorithm.}
\begin{ruledtabular}
\begin{tabular}{ccc}
Temperature & Number of & Number of \\ 
& ``solid'' outcomes & ``liquid'' outcomes\\ \hline
896 & 10 & 0 \\ 
928 & 4 & 6 \\
960 & 3 & 17\\ 
\end{tabular}
\end{ruledtabular}
\end{table}

Following that, the algorithm initiates Gaussian Process (GP) regression (Section \ref{sec:GP}), referred to as the second level. The GP continues to operate based on the optimal sampling strategy outlined in Section \ref{sec:OptimalSampling} while the uncertainty of the MP exceeds the specified threshold. Once the MP uncertainty decreases below the threshold, the GP concludes, and we obtain the values of the MP along with its corresponding uncertainty. We provide a graphical illustration of how our algorithm samples the system size and the temperatures in Appendix \ref{sec:illustration_autonomous}.

In Figure \ref{fig:Al_EAM}, the dependence of the MP of Al EAM \cite{Mendelev} on the number of atoms in the system is illustrated. With the aid of our GP, we are able to extrapolate this dependence to the limit of infinite number of atoms. The value obtained from the GP, along with its corresponding uncertainty, is considered as the estimated ``true'' melting point.

\begin{figure}[ht]
\begin{center}
\includegraphics[width=9.5cm]{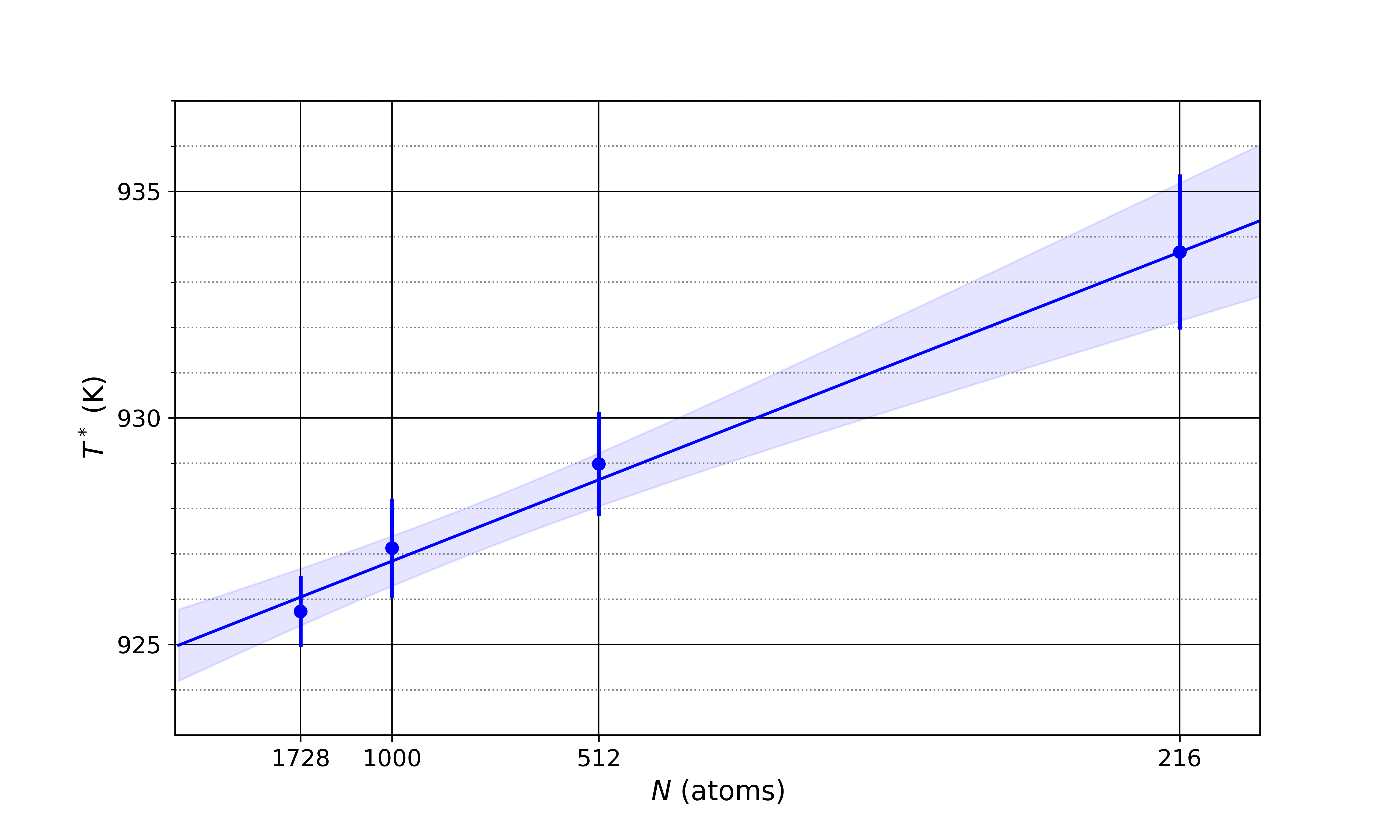}
\caption{Dependence of the melting point $T^*$ on  the number of atoms $N$ for the aluminum EAM \cite{Mendelev}. Points end error bars show values for the fixed number of atoms. The fill shows variance for the ``tested'' number of atoms.}
\label{fig:Al_EAM} 
\end{center}
\end{figure}

\subsection{More examples of unary crystals}

We next present the results of calculations of the melting points for several other systems using different potentials, following a similar approach to what was done for aluminum, as detailed in the previous section. For crystal lattices other than fcc, we ensured that the minimum number of atoms in our simulations was 200. For example, in the case of body-centered cubic (bcc) lattice, we started with $L=4$ (corresponding to 256 atoms), and our initial set for the Gaussian Process (GP) was collected for $L\in\{4,5,6,7\}$. We calculated the melting point for 18 different potentials describing Ag, Al, Cr, Cs, Cu, Fe, K, Li, Na, Ni, Pd, Rb, and Si (some systems were described with more than one potential). The results are summarized in Table \ref{tab:other_examples}.

\begin{table*}
\caption{Melting points (in K) calculated using potentials taken from the following works: \cite{Mendelev,Novikov,AgPd,MEAMFeNiCu,MEAMFe,MEAMNiCu,MEAMCr,Si2017,Alkali}. We used our method to calculate the melting points and compared them to the values obtained in those works. The errors of our results are given as the 1-sigma confidence interval. The underlined results indicate significant deviations (more than 3-sigma) between our values and those published in the other works.}
{
\newcommand{\udash}[1]{%
    \tikz[baseline=(todotted.base)]{
        \node[inner sep=1.2pt,outer sep=0pt] (todotted) {#1};
        \draw[dashed] (todotted.south west) -- (todotted.south east);
    }%
}%
\begin{ruledtabular}
\begin{tabular}{cccccc}
Source  & Material & Potential & Unit cell  & $T_{\text{melt}}$ in this work & $T_{\text{melt}}$ in ref. \\ \hline
Ref. \cite{Mendelev} & Al & EAM & fcc & 925.0 $\pm$ 0.8 & 926 \\ 
Ref. \cite{Novikov}  & Al & MTP & fcc & 887.1 $\pm$ 0.6 & 885\\
Ref. \cite{Mendelev} & Cu & EAM & fcc & 1354.1 $\pm$ 0.7 & 1353\\ 
Ref. \cite{AgPd} & Ag & MTP & fcc & \underline{990.2 $\pm$ 0.6} & \underline{1035}$^{\rm a}$ \\ 
Ref. \cite{AgPd} & Pd & MTP & fcc & \underline{1478.2 $\pm$ 1.0} & \underline{1625}$^{\rm a}$ \\ 
Ref. \cite{MEAMFeNiCu} & Fe & MEAM & bcc & 1809.9 $\pm$ 1.7 & 1812\\
Ref. \cite{MEAMFeNiCu} & Ni & MEAM & fcc & 1704.9 $\pm$ 0.9 & 1705\\ 
Ref. \cite{MEAMFeNiCu} & Cu & MEAM & fcc & 1345.5 $\pm$ 0.7 & 1347\\  
Ref. \cite{MEAMFe} & Fe & MEAM & bcc & \underline{1795.5 $\pm$ 1.3} & \underline{1807}\\  
Ref. \cite{MEAMNiCu} & Cu & MEAM & fcc & \underline{1395.6 $\pm$ 0.7}  & \underline{1320}\\ 
Ref. \cite{MEAMNiCu} & Ni & MEAM & fcc & \underline{1756.1 $\pm$ 1.3} & \underline{ 1742}\\ 
Ref. \cite{MEAMCr} & Cr & MEAM & bcc & \underline{2187.4 $\pm$ 1.8} & \underline{2126} \\  
Ref. \cite{Si2017} & Si & Tersoff & diamond & \underline{1707.3 $\pm$ 0.9} & \underline{1687}\\
Ref. \cite{Alkali} & Li & EAM & bcc & 659.1 $\pm$ 0.6 & 660  \\
Ref. \cite{Alkali} & Na & EAM & bcc & \underline{448.9 $\pm$ 0.9} & \underline{411}  \\
Ref. \cite{Alkali} & K & EAM & bcc & \underline{288.8 $\pm$ 0.9} & \underline{344} \\
Ref. \cite{Alkali} & Rb & EAM & bcc & \udash{329.0 $\pm$ 0.8} & \udash{333} \\
Ref. \cite{Alkali} & Cs & EAM & bcc & \udash{309.0 $\pm$ 0.8} & \udash{305} \\ 

\end{tabular}
\end{ruledtabular}
$^{\rm a}$ 
These values were computed using the Nested Sampling algorithm with a small cell containing 64 atoms. The deviations between the results in \cite{AgPd} and our work can be attributed to the finite-size effect. Confinement to a small cell increases the liquid free energy more significantly than the solid free energy, making the latter more thermodynamically stable at a given temperature.
}
%$^{\rm b}$ These values have been computed using solid-liquid coexistence method, performed in the NPH ensemble, for bcc with approximately 100,000 atoms in a simulation box \cite{MEAMFe} and for fcc with approximately 82,000 atoms in a simulation box. \cite{MEAMNiCu} metals.
%
%$^{\rm c}$ This MP was computed by the interface velocity method.

\label{tab:other_examples}
\end{table*}

We observe that a substantial number of results significantly deviate from our predictions. We consider results to be ``significantly deviating'' when our predictions, taking into account our uncertainty, differ from the literature predictions (to which we assign a 1 K uncertainty), by more than 2 standard deviations. The two results at the bottom exhibit a deviation of approximately 3 standard deviations, while the other underlined results deviate by more than 7 standard deviations. We will discuss the possible reasons for these discrepancies later.

Before discussing the possible reasons for the significant deviations in the results, it is important to highlight that the dependence of the melting point on the system size is not as regular for some of the potentials, unlike what was observed for aluminum as shown in Figure \ref{fig:Al_EAM}.

\begin{figure}[h]
\begin{center}
\includegraphics[width=9.5cm]{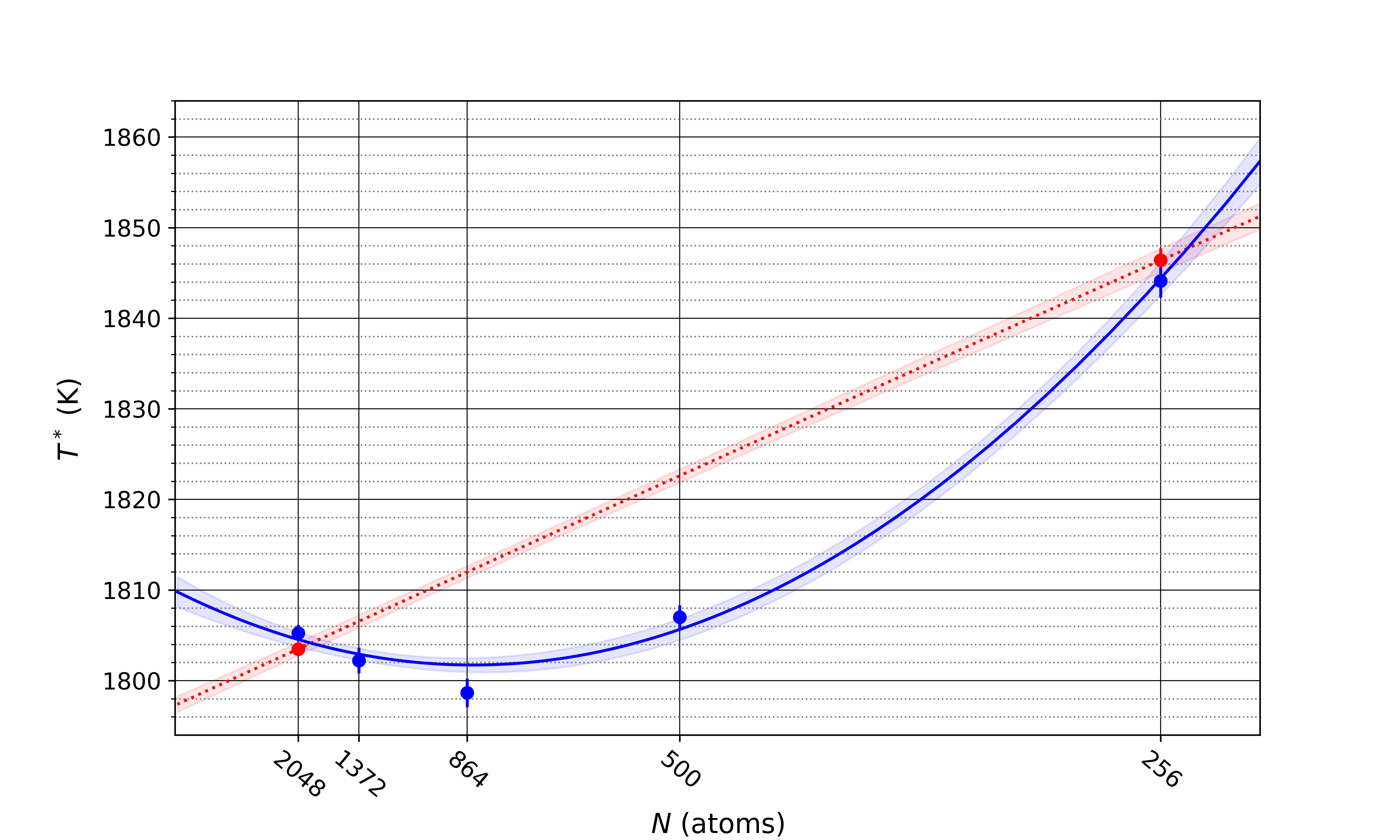}
\caption{
Dependence of the melting point $T^*$ on the number of atoms $N$ for Fe MEAM \cite{MEAMFeNiCu}. The red curve represents the prediction for the melting point if we had only used one system size ($N=256$) as the initial guess for the GP.
The GP in this case adds one more system size ($N=2048$) and makes an inaccurate extrapolation as $N\to\infty$.
The blue curve, on the other hand, demonstrates how the melting point dependence fluctuates as the number of atoms in the system increases, dropping down and then rising again.
In the latter case the GP's prediction coincides with that of \cite{MEAMFeNiCu}.
}
\label{fig:anomalous_Fe} 
\end{center}
\end{figure}

Upon examining the dependencies for the 18 potentials, we identified four MEAM potentials (Fe \cite{MEAMFeNiCu}, Fe \cite{MEAMFe}, Ni \cite{MEAMNiCu}, and Cr \cite{MEAMCr}) whose dependence on the system size is irregular. An example of such irregular dependence is shown in Figure \ref{fig:anomalous_Fe} for the Fe potential from \cite{MEAMFeNiCu}. It can be observed that as the system size increases, the melting point initially decreases to around 1800 K but then rises to 1810 K.

From these observations, we draw several conclusions. Firstly, if we had used only one system size as the initial data for our GP, it would have determined that one extra point is sufficient. Based on these two points, the GP makes a prediction that is below 1800 K. This is why we enforce the collection of data from at least four different system sizes before making the final prediction (the solid line in Figure \ref{fig:anomalous_Fe} has five points, with the fifth point added by our automatic sampling algorithm).

Secondly, although our algorithm should, in theory, predict the melting point accurately in the limit of infinite data, the behavior and trend of the data as the system size increases, as well as the uncertainty of the prediction, may not be accurately identified for finite data. This suggests that the kernel \eqref{eq:kernel} may not accurately describe the data. %A detailed investigation of this issue is deferred to future work.

Thirdly, the fact that only MEAM potentials exhibit such irregular behavior raises the question of whether this phenomenon has a physical basis or if it is an artifact of the functional form of MEAM itself. The absence of such behavior in the three machine-learning potentials (for Al, Ag, and Pd), which have richer functional forms than MEAM, as well as in the EAM potentials with a more restrictive functional form, suggests the possibility that it may be an artifact of the MEAM functional form.

Further exploration and analysis of these observations will be conducted in future research.

\begin{figure}[h]
\begin{center}
\includegraphics[width=9.5cm]{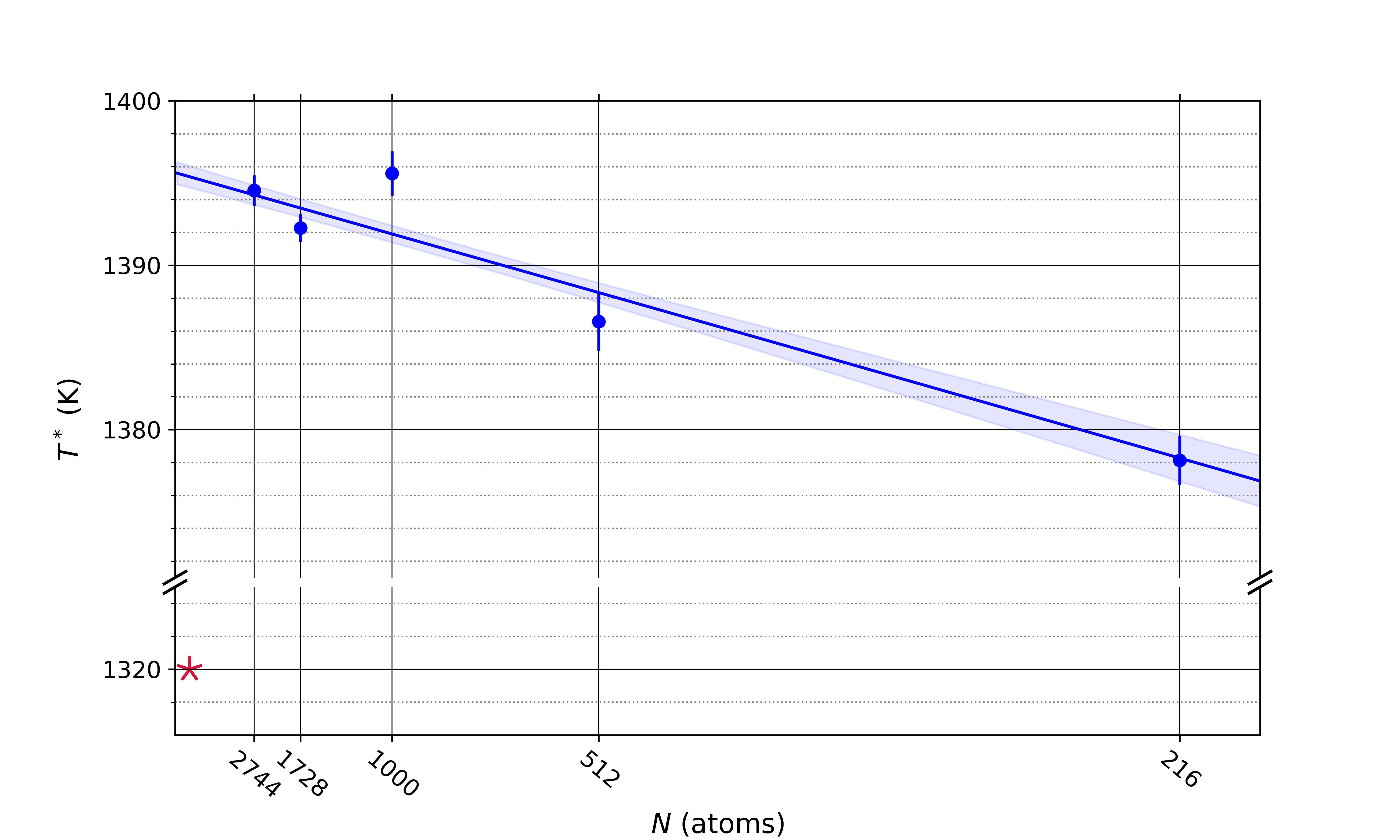}
\caption{Dependence of the melting point $T^*$ on the number of atoms $N$ for Cu MEAM \cite{MEAMNiCu}. The blue curve represents our results, while the red star represents the result from \cite{MEAMNiCu}, which is nearly 80 K lower than our prediction in the limit of an infinite number of atoms. 
}
\label{fig:anomalous_Cu} 
\end{center}
\end{figure}

We now turn to the issue of large deviations, by more than 2-sigma, between our results and those in the literature as presented in Table \ref{tab:other_examples}. The 3-sigma deviations \emph{may} be explained by the irregular behavior of the melting point on the system size. Additionally, the deviations for Ag and Pd from \cite{AgPd} are explained by the fact that MPs were computed in \cite{AgPd} using the nested sampling algorithm \cite{NS1,NS2,NS3} with 64 atoms, and the finite-size error can easily exceed 100 K in this case.
However, there are still seven results in the table with deviations exceeding 7-sigma, and we believe that they cannot be
%\hl{fully}
explained by shortcomings of either our method or the methodologies applied in the referenced works. 
%\hl{One possible source of error in the referenced works can, nevertheless, be conjectured based on our work, namely that the equilibration time of the NPH-coexistence simulations should be of the same scale as the NPT case---$O(L^4)$ or $O(N^{4/3})$ (see Section \ref{sec:simulation_time}.)
%This means that, if we take 10-100 ps as a typical time scale for the solid-liquid interface to drift by a considerable fraction of the supercell, as suggested by Figure \ref{fig:lines_time_en}, the NPH simulations with about $10^5$ atoms \cite{MEAMFe,MEAMNiCu} would take about 2--20 ns to equilibrate, whereas they are typically equilibrated over 100--500 ps \cite{MEAMFe,MEAMNiCu}.
%}
As an example, we plot the MP versus system size dependence for Cu from \cite{MEAMNiCu} in Figure \ref{fig:anomalous_Cu}. Indeed, by looking at our results, it is hard to conjecture the cause of the error assuming that the true melting point is 80 K lower than our prediction. Moreover, the NPH-coexistence method is well-validated, and many other results obtained using this method coincide with our results within the 2-sigma interval.
Interestingly, we are not in a typical situation where the results of two conventionally performed studies, which used manual protocols of preparation and well-established methods, differ significantly. In such cases, one would expect that at least one study made a mistake in manually handling the data. In our case, we are comparing a conventionally performed study with the results obtained from an autonomously functioning system, which, although not without its own shortcomings as mentioned earlier, largely makes decisions autonomously (albeit within predetermined parameters) and, in theory, should produce exact results given infinite data.

\subsection{Examples of binary compounds}

We will now discuss and demonstrate the application of the proposed method to binary systems. To begin, we recall that one of the motivations for our work is the fact that thermodynamic integration can accurately reconstruct the free energy of solid and liquid phases up to additive shifts. The melting point serves as a crucial datum from which this relative additive shift between the two phases can be determined. Our specific motivation comes from the work by Ladygin et al. \cite{Ladygin}, where a semi-automatic algorithm for constructing phase diagrams from MD calculations was proposed.

\begin{figure}[b]
\includegraphics[width=8cm]{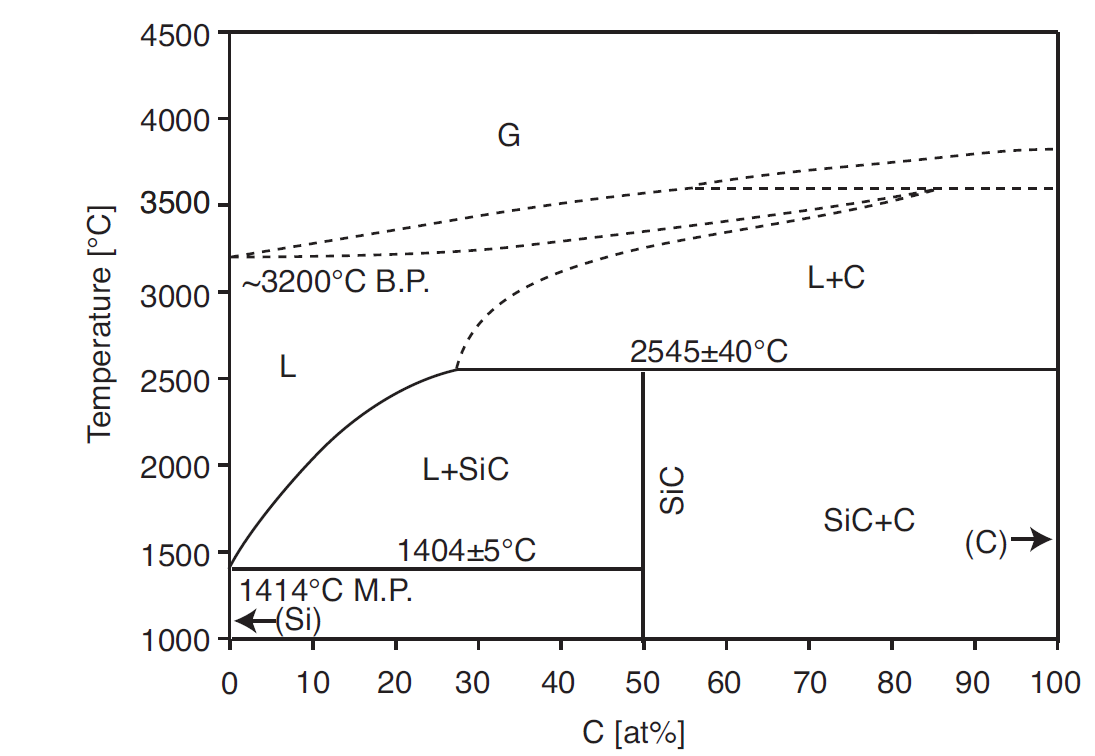} \\
\caption{\label{fig:si-c} The Si-C phase diagram \cite{nishitani2011metastable,olesinski1996b}.
}
\end{figure}

Let us consider Silicon carbide (SiC) as an example. The Si-C phase diagram (Figure \ref{fig:si-c}) consists of three solid phases: Si, SiC, and C, along with one fluid phase. Our method is directly applicable to the unary phases. However, the melting point of SiC corresponds to the triple point of liquid, SiC, and C coexistence. The strategy we advocate is therefore to obtain the melting point of Si, as we did in the previous subsection. Then, by using thermodynamic integration, we can reconstruct the free energy profiles of Si, SiC, C, and liquid, ultimately enabling the determination of the melting point of SiC.
Phonon calculations could be used to fix this additive constant for C and SiC, while the melting point of pure Si, which can be found robustly with out algorithm, can be used to find the additive constant for the liquid.

However, for the purpose of demonstration, we apply our method to SiC, utilizing the interatomic potential from \cite{kang2014-SiC}. In that study, the authors estimated the melting point to be 2600 K.
If we apply our algorithm directly, the coexistence simulations run indefinitely while attempting to determine the resulting phase.
If we choose to adjust the solid and liquid threshold coefficient from 1/8 to 1/4 then the coexistence simulations terminate, however, they result in incorrect phases. Indeed, in the simulations, the solidified phase resembles a random solution, exhibiting a higher enthalpy than the solid phase. Furthermore, the phase diagram suggests coexistence of graphite and liquid Si-C, but our simulation shows a pure liquid phase.
We emphasize that changing the threshold from 1/8 to 1/4 does not ``fix'' the algorithm, but merely allows us to investigate the reasons behind its failure.
Consequently, our simulations indicate the temperature of coexistence for unstable phases, and the resulting melting point should not be considered quantitatively accurate. Instead, a more robust approach, as described above, should be employed.

\begin{table*}
\caption{Melting points (in K) for MgO and CaO calculated using potentials from \cite{MgCaO}. We used our method to calculate the melting points and compared them to the values obtained in those works. The errors of our results are given as the 1-sigma confidence interval.
In \cite{MgCaO} two methods were used: thermodynamic integration (TI) and NpH coexistence simulations.
The two underlined coexistence results indicate significant deviations, by about 6-sigma between our values and those published in the other works. Because of large uncertainty of the TI values in \cite{MgCaO}, the deviations of those values with our results are not statistically significant.}
\begin{ruledtabular}
\begin{tabular}{cccccc}
Source  & Material & Potential & Unit cell  & $T_{\text{melt}}$ in this work & $T_{\text{melt}}$ in ref. \\ \hline
\multirow{2}{*}{Ref.\ \cite{MgCaO}} & \multirow{2}{*}{MgO} & \multirow{2}{*}{PBE-NNP} & \multirow{2}{*}{cubic} & \multirow{2}{*}{2800.5 $\pm$ 1.0} & \underline{2786 $\pm$ 1.5} (coexist) \\
& & & & & 2787 $\pm$ 30 (TI) \\[3pt]

\multirow{2}{*}{Ref.\ \cite{MgCaO}} & \multirow{2}{*}{MgO} & \multirow{2}{*}{SCAN-NNP} & \multirow{2}{*}{cubic} & \multirow{2}{*}{3184.9 $\pm$ 1.4} & 3181 $\pm$ 1.7 (coexist) \\
& & & & & 3173 $\pm$ 33 (TI) \\[3pt]

\multirow{2}{*}{Ref.\ \cite{MgCaO}} & \multirow{2}{*}{CaO} & \multirow{2}{*}{PBE-NNP} & \multirow{2}{*}{cubic} & \multirow{2}{*}{2650.5 $\pm$ 1.3} & 2659 $\pm$ 1.5 (coexist) \\
& & & & & 2640 $\pm$ 30 (TI) \\[3pt]

\multirow{2}{*}{Ref.\ \cite{MgCaO}} & \multirow{2}{*}{CaO} & \multirow{2}{*}{SCAN-NNP} & \multirow{2}{*}{cubic} & \multirow{2}{*}{3082.3 $\pm$ 1.0} & \underline{3097 $\pm$ 2.0} (coexist) \\
& & & & & 3057 $\pm$ 35 (TI) \\
\end{tabular}
\end{ruledtabular}
\label{tab:binary_examples}
\end{table*}

Nevertheless, if coexistence simulations involve only two stable phases competing with each other, our algorithm directly applies to the case of binary compounds. We applied our algorithm to calculate the melting points of MgO and CaO using the neural-network potentials from \cite{MgCaO}. Our results exhibit a relatively good agreement with \cite{MgCaO}. The thermodynamic integration results of \cite{MgCaO}, despite their significant uncertainty of about 30 K, do not exhibit any statistically significant disagreement with our results. However, the coexistence simulation results of \cite{MgCaO} deviate from our results by about 15 K, which is statistically significant, although small compared to the absolute values of the melting points, around 3000 K. The disparity between our results and the literature emphasizes the need for robust and error-proof methods for melting point calculations.

\section{Conclusion}\label{sec:conclusions}

We have developed an algorithm for computing the melting point given the interatomic interaction potential of an element and its crystal structure by autonomously learning from coexistence simulations. Our algorithm makes optimal decisions about the number of atoms $N$ and temperature $T$ at which to conduct the NPT coexistence simulations, collects data from these simulations, and constructs a machine-learning model. This model enables predictions of the melting point in the limit as $N$ approaches infinity. To design our method, we rely on physical models. We solve the Fokker-Planck equations that describe the motion of the solid-liquid interface, allowing us to derive a nonlinear Bayesian regression likelihood formula for the melting point. The scaling of the model coefficients and solution with $N$ facilitates the determination of the model hyperparameters. In theory, with infinite data, the algorithm should converge to the true melting point. However, we observed that at intermediate stages (where we used our algorithm to predict the melting point with an error of approximately 1 K), the algorithm appears to somewhat underestimate the predictive error. Nonetheless, we have applied our algorithm to approximately 20 interatomic potentials for various materials. In about one-third of these cases, we observed significant deviations from the results published in the literature, which emphasizes the need for automatic and reliable algorithms for melting point calculations.

\section{Acknowledgments}\label{sec:acknowledgments}

This work was supported by Russian Science Foundation (grant number 23-13-00332, https://rscf.ru/project/23-13-00332/).

\bibliography{refs}

\appendix

\section{Solution to the Fokker-Planck equation}

In order to derive  the probability of reaching for instance, the all-solid state \eqref{eq:p_sol_formula} from \eqref{eq:FP_eq} and \eqref{eq:FP_bc}, we formulate the following stationary model:
\begin{equation*}
    D p''+ \alpha p' = \delta (x)
\end{equation*}
in terms of $p=p(x)$ with the boundary conditions
\begin{equation*}
    p(-L) = p(L) = 0.
\end{equation*}
This model describes an ensemble of trajectories that are continuously generated at $x=0$ --- hence the Dirac delta $\delta(x)$ --- and terminated at $x=\pm L$.
By solving the above system analytically, we can derive the probability of the trajectory reaching the all-solid state $x=L$ as
\begin{align*}
p_\sol &= \frac{-p'(L)}{p'(-L)-p'(L)} = \frac{1}{1 + e^{\gamma(T-T^*) L^3/\kappa}}\nonumber\\ 
&= \frac{1}{1 + e^{(T-T^*)/\sigma^*}},
\end{align*}
which is the same as \eqref{eq:p_sol_formula}.
Here we define $\sigma^* = \kappa/(\gamma L^3)$ which we call the \emph{melting temperature spread}.
    
\section{Dependence of the melting point of Al on the number of k points}

We apply our methodology to investigate the correlation between the MP temperatures and the convergence parameters of the Density Functional Theory (DFT) used in the computation of those melting points (MPs). 
This is achieved by employing MTP (Machine-Learning Potential) and fitting it to the DFT data in a fully automatic manner, actively learning from DFT while performing MD simulations \cite{Novikov}.
Specifically, we focus on the case of aluminum, where we fit eight potentials to DFT calculations. The first potential is fitted based on a single gamma point, followed by potentials fitted on $2\times2\times2$, $3\times3\times3$ k-point meshes, and so on.
We used the GGA-PBE density functional as implemented in the VASP software \cite{VASP1,VASP3,VASP4}.
The GGA-PBE functional is recognized for its tendency to underestimate the strength of interatomic bonds, resulting in lower melting points.
A PAW pseudopotential with three valence electrons was used to model aluminum.
ENCUT is set to 410 eV, which is 1.7 times the ENMAX and ensures negligible errors in energy differences (at least when compared to the training errors).
The training errors and the results of the MP calculations are presented in Table \ref{tab:al-dft}.
It is noteworthy that while the fitting errors reach a plateau as the k-point mesh is refined starting from the $4\times4\times4$ mesh, the melting point continues to increase, ranging from 894 K to the 903.4--905 range for the more refined meshes.

\begin{table}[h]
\caption{The melting points (in K) calculated using eight MTPs \cite{shapeev_moment_2016} fitted to DFT calculations performed with different k-point meshes. The corresponding fitting errors are also provided.
}\label{tab:al-dft}
\begin{ruledtabular}
\begin{tabular}{cccc}
kpts & $T_{\text{melt}}$ (K) & en.error & force error \\ 
& & (meV/atom) & (meV/\AA) \\ \hline
$1\times1\times1$   & 880.3 $\pm$ 1.1 & 2.748 & 108.8 \\
$2\times2\times2$   & 977.2 $\pm$ 1.1 & 1.705 & 39.0 \\ 
$3\times3\times3$   & 886.9 $\pm$ 1.0 & 0.669 & 29.4 \\ 
$4\times4\times4$   & 894.1 $\pm$ 1.1 & 0.670 & 26.3 \\
$5\times5\times5$   & 905.0 $\pm$ 1.1 & 0.650 & 26.7 \\ 
$6\times6\times6$   & 902.3 $\pm$ 1.1 & 0.630 & 25.4 \\ 
$7\times7\times7$   & 900.7 $\pm$ 1.0 & 0.646 & 25.2 \\ 
$8\times8\times8$   & 903.4 $\pm$ 1.0 & 0.644 & 25.3 \\ 
\end{tabular}
\end{ruledtabular}
\end{table}

It is also interesting to compare these results with other machine-learned potentials as well as those obtained from accurate DFT calculations.
In \cite{Novikov}, the authors used MTPs and an NVE coexistence method to determine a melting point of 885 K for Al on a $3\times3\times3$ k-point mesh, which is in agreement with our present results.
Accurate DFT computations with an $8\times8\times8$ k-point mesh resulted in a melting point of 888 K \cite{zhu2020-al-dft}. Considering the limited sampling possible with such expensive DFT calculations, this value agrees well with our result of 900--905 K.
In \cite{pun2020-pinn}, the authors trained a PINN model on a $2\times2\times2$ k-point mesh and obtained a melting point of 975 $\pm$ 3 K, which again closely coincides with our calculations. In \cite{smith2021-ani-al}, on the other hand, the authors fitted a potential for Al and obtained a melting point of 925 K, which is different from the values obtained in \cite{Novikov,pun2020-pinn,zhu2020-al-dft} or this work.
%The discrepancy between this work, \cite{Novikov}, and \cite{pun2020-pinn} on one hand, and \cite{smith2021-ani-al} on the other hand, as we believe, is unlikely to be caused by an error in \cite{smith2021-ani-al}, but rather a fortunate coincidence that the potential fitted in \cite{smith2021-ani-al} better reproduces the experimental result (933 K) than the underlying DFT data \cite{smith2021-personal}.
%It is important to note that potentials do not always fit all properties equally accurately, and often the potential that better reproduces experimental values is chosen if multiple potentials were fitted.

\section{Illustration of autonomous learning}\label{sec:illustration_autonomous}

\begin{figure}
    \begin{tabular}{c}
     \includegraphics[width=7.5cm]{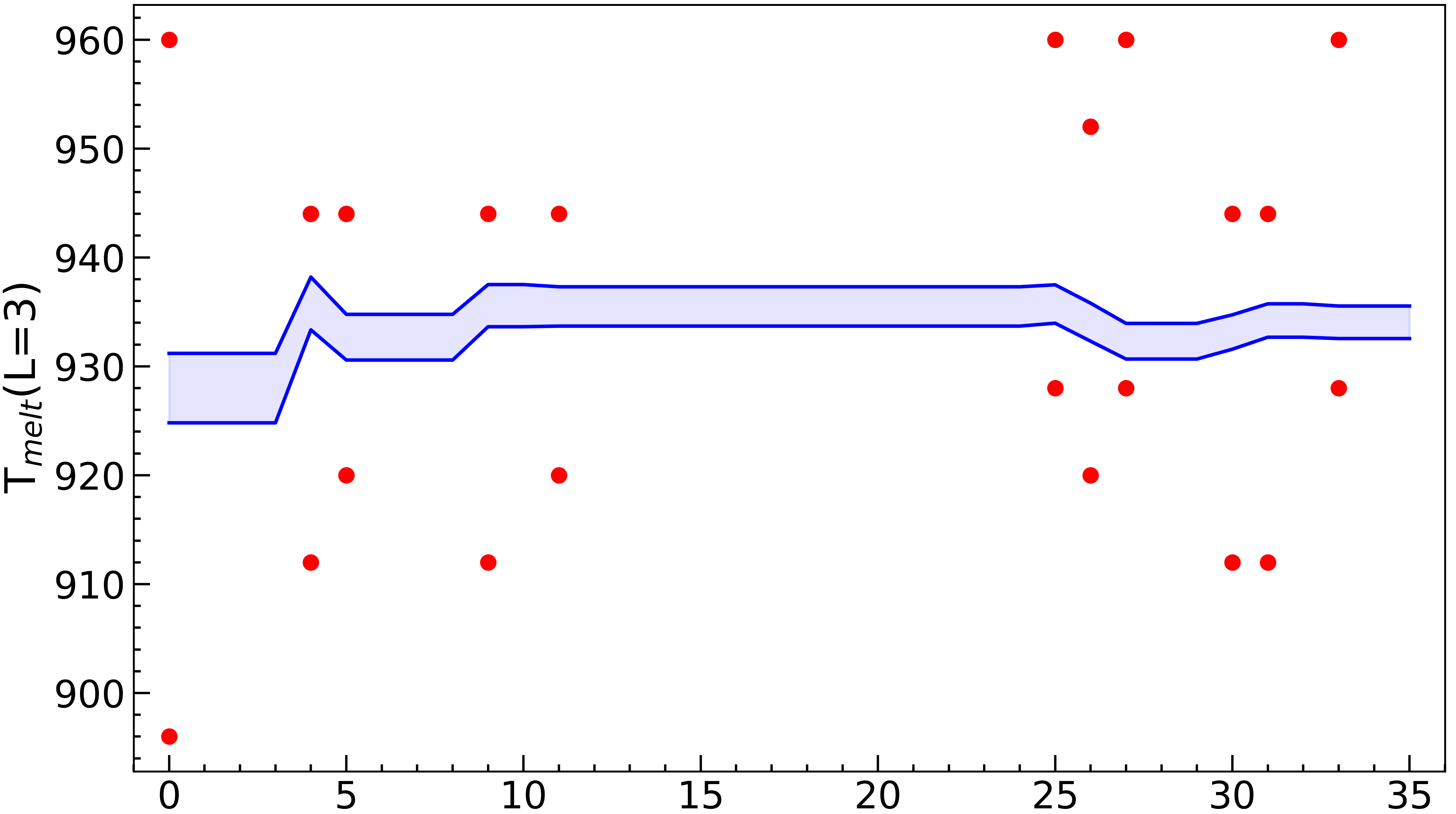}\\
     \includegraphics[width=7.5cm]{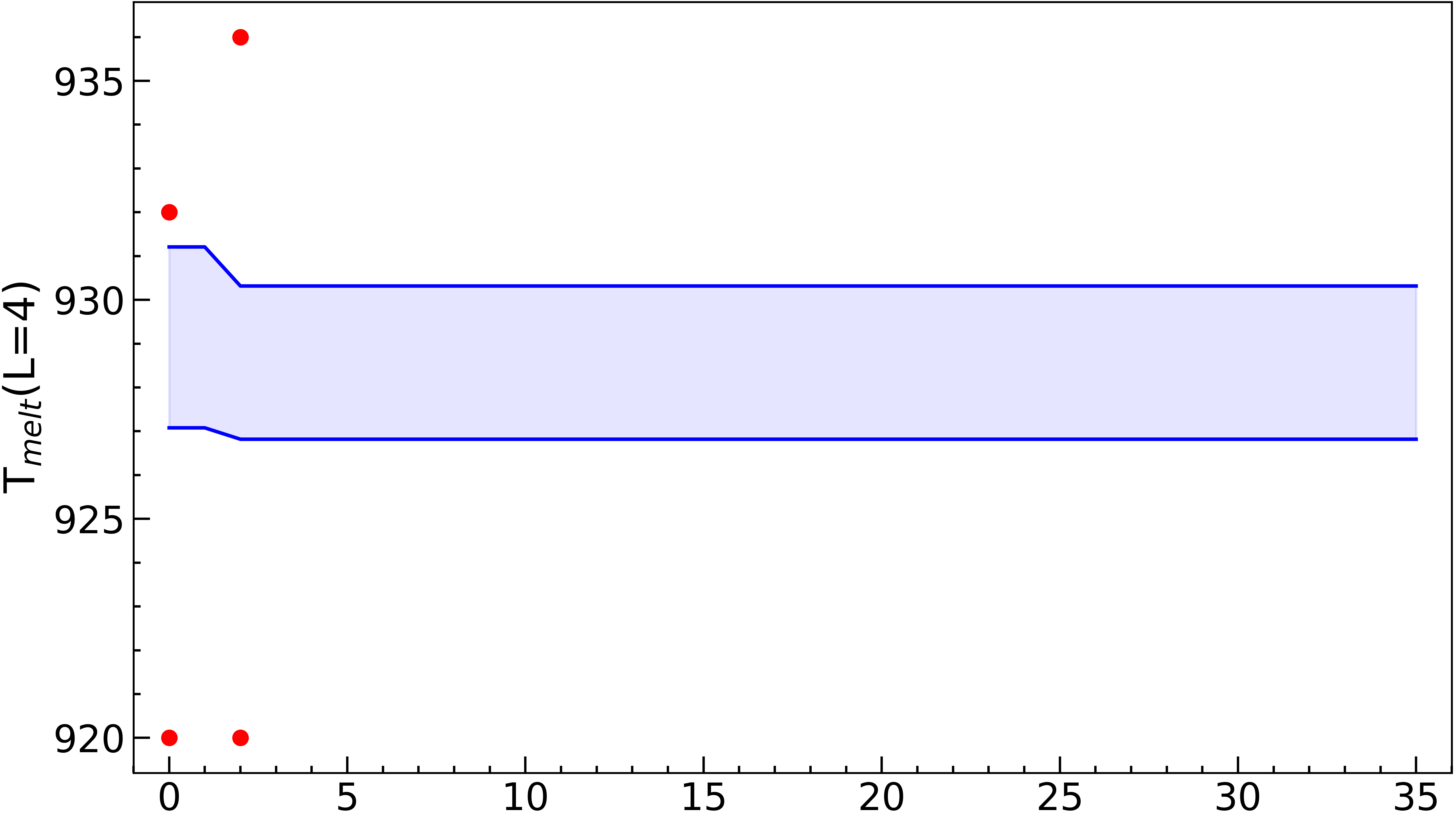}\\ 
     \includegraphics[width=7.5cm]{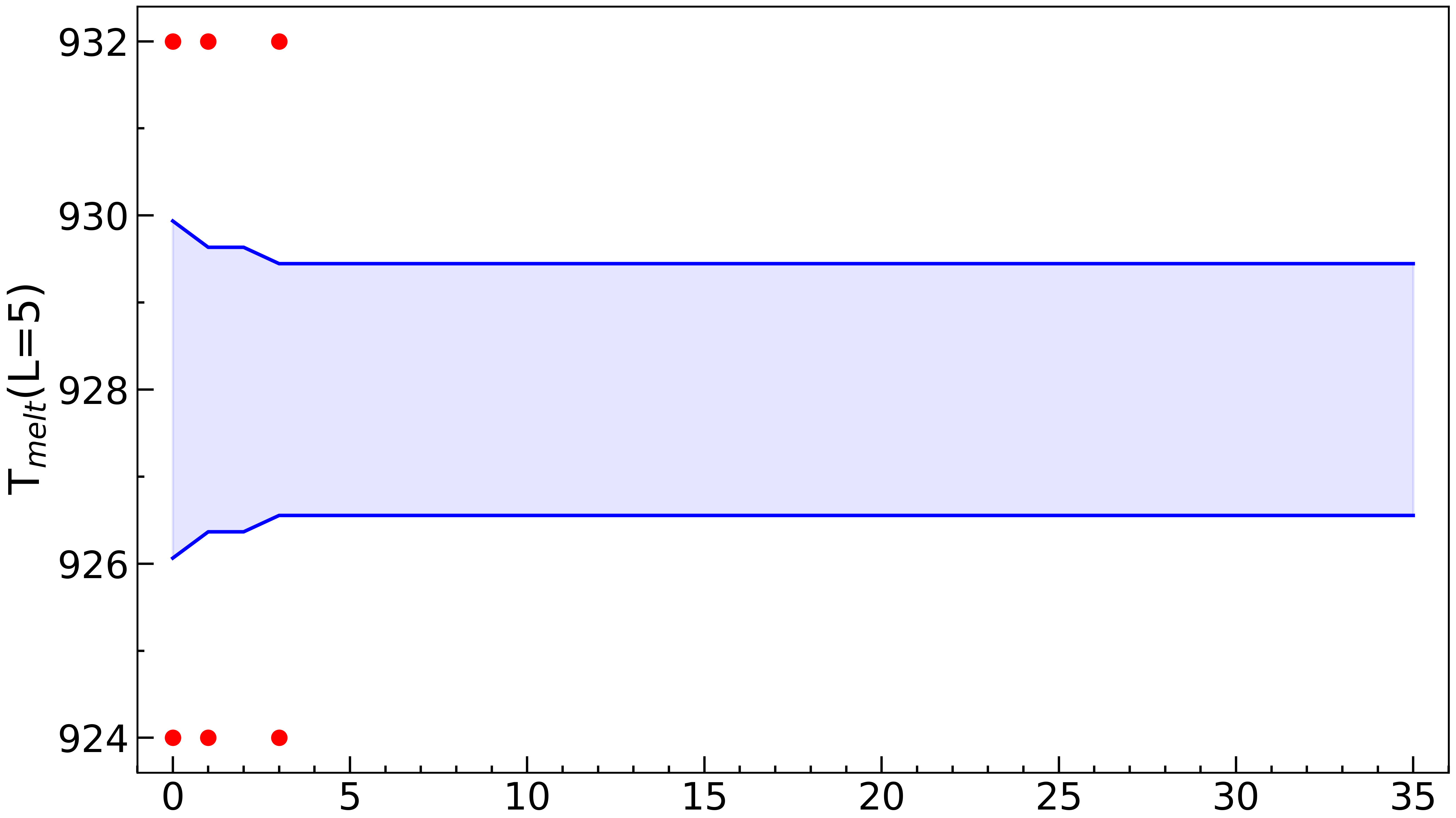}\\ 
     \includegraphics[width=7.5cm]{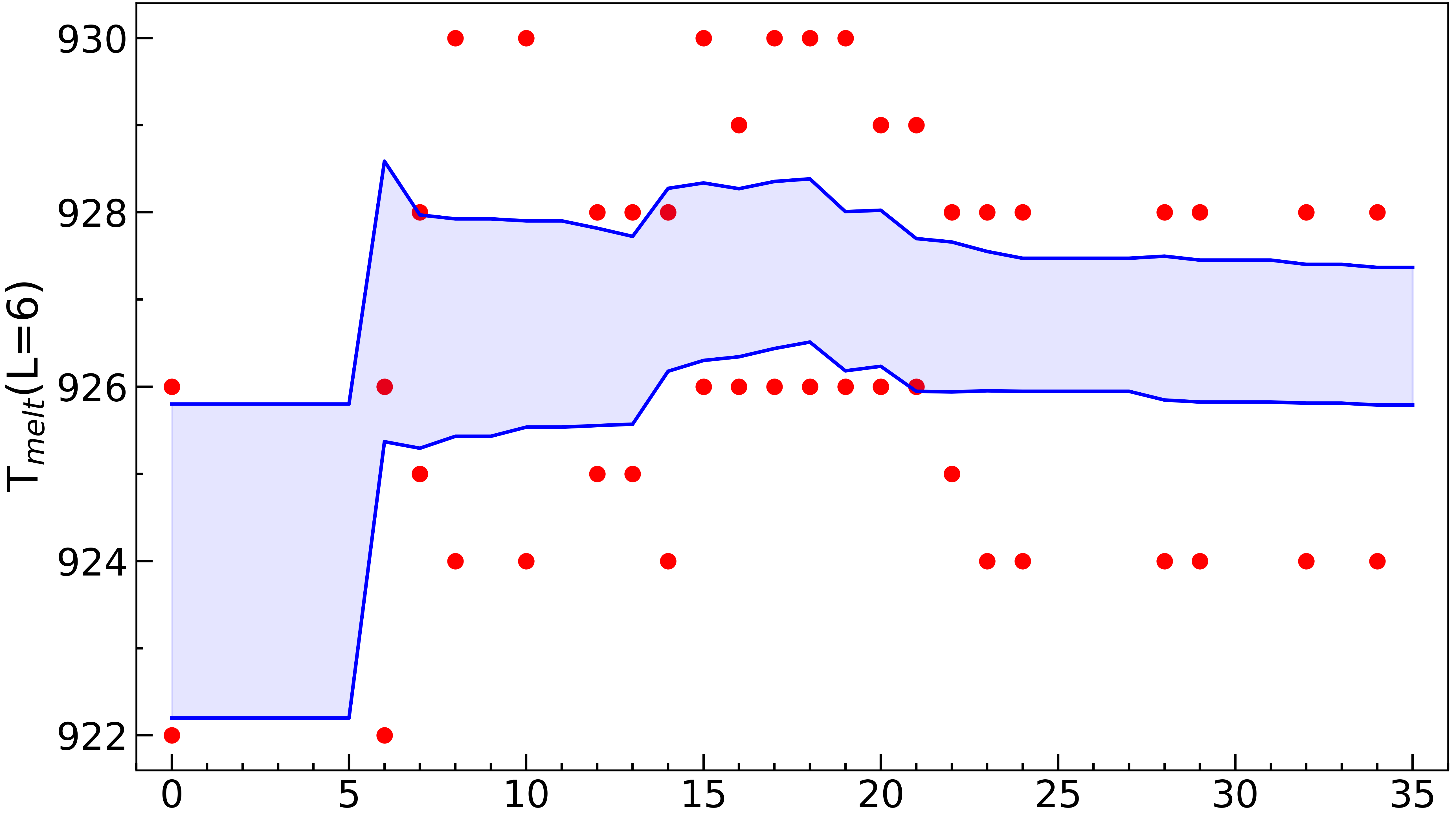}\\
     \includegraphics[width=7.5cm]{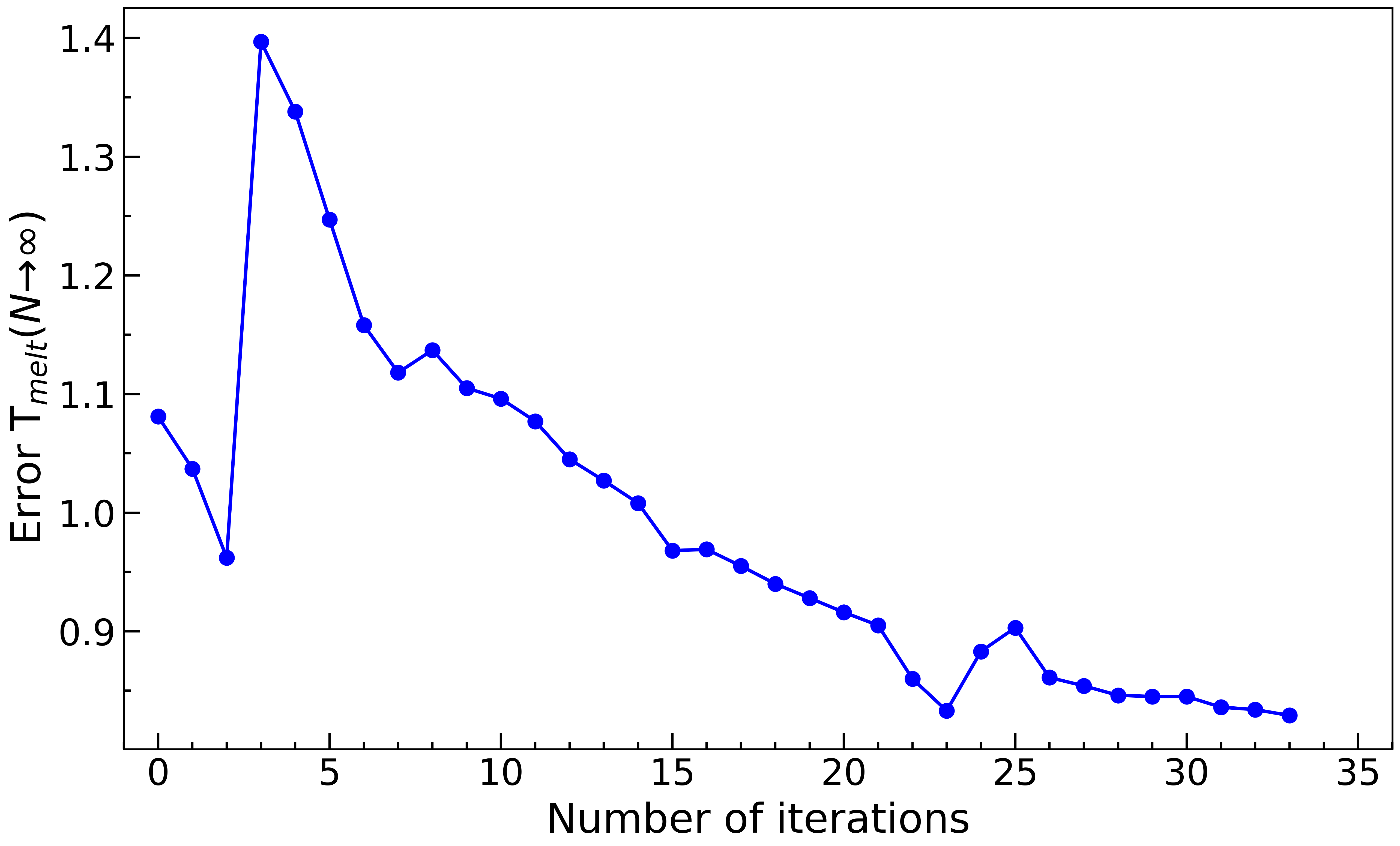}\\
    \end{tabular}
    \caption{Illustration of how our algorithm chooses the system size $L$ and temperature $T$ with which to conduct simulations; the details are in the text.
    }
    \label{fig:data_for_each_L}
\end{figure}

The process by which our algorithm selects the system size $L$ and temperature $T$ for conducting simulations is illustrated in Figure \ref{fig:data_for_each_L}.
The algorithm starts with data computed for $L\in\{3,4,5,6\}$ and two temperatures for each $L$.
The four plots in Figure \ref{fig:data_for_each_L} correspond to these four values of $L$. Subsequently, the algorithm proceeds through 34 iterations (the iteration number is represented on the X-axis). During each iteration, it selects a specific $L$ and conducts simulations with two different temperature values. The temperatures chosen for each iteration are plotted as red points on the graph corresponding to the selected $L$. The cumulative results of these simulations reduce the uncertainty associated with the respective $L$ (as shown by the shaded region) and, consequently, decrease the error in extrapolation as $L\to\infty$ (as depicted in the fifth graph). We can see that, according to the algorithm, it is most efficient to choose either $L=3$ or $L=6$ to minimize the error.

\end{document}